\documentclass[12pt]{article}
\pdfoutput=1

\usepackage{putex}
\usepackage{graphicx}
\usepackage{caption}
\usepackage{ dsfont }
\usepackage{amsmath}
\usepackage{array}
\usepackage{subcaption}
\usepackage{epstopdf}
\usepackage{enumerate}
\usepackage{cite}
\usepackage{youngtab}
\usepackage{tensor}
\usepackage[normalem]{ulem}
\usepackage{slashed}
\usepackage[aligntableaux=center]{ytableau}
\usepackage[utf8]{inputenc}
\usepackage[
      colorlinks=true,
      linkcolor=blue,
      urlcolor=blue,
      filecolor=black,
      citecolor=red,
      ]{hyperref}
\usepackage{braket}
\usepackage{mathtools}
\usepackage{rotating}
\usepackage{tabularx}
\newcolumntype{Y}{>{\centering\arraybackslash}X}
\newcolumntype{C}[1]{>{\centering\arraybackslash}p{#1}}
\usepackage{todonotes}
\usepackage{color, colortbl}
\definecolor{LightCyan}{rgb}{0.7,1,1}
\definecolor{Gray}{gray}{0.9}
\usepackage{xspace}

\newcommand{\abs}[1]{\left\lvert #1 \right\rvert}

\newcommand {\be} {\begin {equation}}
\newcommand {\ee} {\end {equation}}

\newcommand {\bes} {\begin {equation*}}
\newcommand {\ees} {\end {equation*}}

\newcommand{\es}[2] {\begin{equation} \label{#1} \begin{split} #2 \end{split} \end{equation}}

\newcommand{\cA}{{\mathcal A}}

\newcommand{\cF}{{\mathcal F}}

\newcommand{\cK}{{\mathcal K}}
\newcommand{\cL}{{\mathcal L}}
\newcommand{\cN}{{\mathcal N}}
\newcommand{\cO}{{\mathcal O}}

\newcommand{\cR}{{\mathcal R}}

\newcommand{\cM}{{\mathcal M}}

\newcommand{\beq}{\begin{equation}}
\newcommand{\eeq}{\end{equation}}

\def\p{\partial}

\def\ie{\begin{equation}\begin{aligned}}
\def\fe{\end{aligned}\end{equation}}

\newcommand{\m}{\mu}
\newcommand{\n}{\nu}
\newcommand{\pa}{\nabla}

\newcommand{\mf}{\mathfrak }

 \newcommand{\zzb}{\mathcal{X}}

\numberwithin{equation}{section}

\def\<{\langle}
\def\>{\rangle}

\newcommand{\ak}{\alpha}        
\newcommand{\bk}{\beta}         
\newcommand{\gk}{\gamma}        
\newcommand{\dk}{\delta}        \newcommand{\Dk}{\Delta}

\newcommand{\lk}{\lambda}       
          
\newcommand{\sk}{\sigma}        
\newcommand{\tk}{\tau}

\def\pa{\partial}

\def\btau{{\bar\tau}}

\DeclareMathOperator{\Tr}{Tr}

\newcommand{\Kahler}{K\"{a}hler\xspace}
\newcommand{\bbeta}{{\bar\beta}}
\newcommand{\bZ}{\bar{Z}}
\newcommand{\bX}{\bar{X}}
\newcommand{\bW}{\bar{W}}
\newcommand{\bz}{\bar{z}}

\def\beg{\begin{equation}\begin{gathered}}
\def\eeg{\end{gathered}\end{equation}}

\def\ttk{\tilde\tk}

\def\bW{{\bar W}}
\def\bb{{\bar\bk}}
\def\bd{{\bar\dk}}

\preprint{PUPT-2628  \\ MIT-CTP/5379}

\institution{PU}{Joseph Henry Laboratories, Princeton University, Princeton, NJ 08544, USA}
\institution{Stanford}{Stanford Institute for Theoretical Physics, Department of Physics, \cr Stanford University,  Stanford, CA 94305, USA}
\institution{MIT}{Center for Theoretical Physics and Department of Mathematics,  \cr Massachusetts Institute of Technology, Cambridge, MA 02139, USA}

\title{
The $\cN = 2$ Prepotential and the Sphere Free Energy
}

\authors{Bernardo Zan,\worksat{\PU} Daniel Z.~Freedman,\worksat{\Stanford, \MIT} and Silviu S.~Pufu\worksat{\PU}  }

\begin{document}

\abstract{
We study the mass-deformed sphere free energy of three-dimensional ${\cal N} = 2$ superconformal field theories with holographic duals.  Building on previous observations, we conjecture a proportionality relation between the sphere free energy on the boundary and the prepotential of the four-dimensional ${\cal N} = 2$ supergravity theory in the bulk.  We verify this formula by explicit computation in several examples of supergravity theories with vector multiplets and hypermultiplets. %
}
\date{December 2021}

\maketitle

\tableofcontents

\section{Introduction}
Three-dimensional ${\cal N} = 2$ superconformal field theories with flavor symmetry admit supersymmetric ``real mass'' deformations \cite{Aharony:1997bx}.  As explained in \cite{Aharony:1997bx}, the real mass parameters, denoted here generically by $\mathfrak{m}$, are introduced by first coupling the conserved current multiplet to a background vector multiplet, and then giving the vector multiplet scalars supersymmetry-preserving expectation values proportional to $\mathfrak{m}$.  These real mass deformations can be constructed not only in flat space, but also on curved manifolds, and in this work we will be interested in the case of the round sphere $S^3$.  As explained in \cite{Jafferis:2010un,Hama:2010av}, the mass-deformed $S^3$ free energy $F_{S^3}(\mathfrak{m})$ can be computed exactly in any Lagrangian ${\cal N} = 2$ theory using the technique of supersymmetric localization (see \cite{Willett:2016adv,Pufu:2016zxm,Dumitrescu:2016ltq} for reviews and references).  Although we will not make use of this fact explicitly, we should point out that the mass-deformed sphere free energy is related by analytic continuation to the trial R-charge deformations used in the $F$-maximization procedure \cite{Jafferis:2010un,Closset:2012vg}, so we will not make a distinction between trial R-charges and mass parameters in the following discussion.

The focus of our work is to study $F_{S^3}(\mathfrak{m})$ in holographic theories at leading order in the bulk derivative expansion.  In recent work \cite{Binder:2021euo}, it was pointed out that, in a holographic setup, $F_{S^3}(\mathfrak{m})$ receives contributions only from a restricted class of supersymmetric terms in the bulk effective action in $AdS_4$.  In the terminology introduced in  \cite{Binder:2021euo}, the bulk deformations that affect the sphere free energy are chiral F-terms and flavor current terms.  On the other hand, $F_{S^3}(\mathfrak{m})$ is independent of bulk D-terms, $1/4$-BPS terms, and non-chiral F-terms.  At the two-derivative level in the bulk effective action, the only chiral F-terms are prepotential interactions, so in theories without flavor current terms, one concludes that the mass-deformed sphere free energy should be completely determined by the prepotential of the bulk supergravity theory on $AdS_4$.

In this paper we study $SO(4)$-invariant %
``domain wall'' solutions of ${\cal N} = 2$ supergravity theories with asymptotically $AdS_4$ metrics, which are putatively dual to 3d ${\cal N} = 2$ superconformal field theories (SCFTs) on $S^3$ deformed by the mass parameters $\mathfrak{m}$.  We conjecture a simple formula relating the classical prepotential of the theory to its mass-deformed free energy $F_{S^3}(\mathfrak{m})$.  Since the free energy can be viewed as an observable of the bulk theory, this formula should follow from the structure of the bulk Lagrangian, whether or not a theory has an AdS/CFT dual.  The conjectured
formula is verified in several explicit models with only vector multiplets, and in models with an added hypermultiplet.  
We next introduce the essential machinery of vector multiplet models needed to present our conjecture with details
to be further explained in Section~\ref{sec: abelian vector mults}.

\subsection{The conjecture}

The Lagrangian of a model containing $n_V$ vector multiplets  is determined by its prepotential ${\cal F}(X^I)$, a homogeneous function of degree two, i.e.~${\cal F}(\lambda X^I) =\lk^2 {\cal F}(X^I)$.  The $X^I$, with $I=0,1,\dots n_V$, are holomorphic projective coordinates of a special \Kahler manifold of complex dimension $n_V$. The $X^I$ are related to physical scalar fields $\tau^\ak,\,\,\ak= 1,2,\ldots, n_V$,  which provide intrinsic coordinates for this manifold.  %
In Lorentzian signature, the fields $\tau^\alpha$ have $\bar \tau^{\bar \alpha}$ as their complex conjugates;  however, in Euclidean signature the two sets of fields are independent, and in order to emphasize this fact we use the notation $\tilde \tau^{\tilde \alpha}$ instead of $\bar \tau^{\bar \alpha}$.  The \Kahler potential $\cK(\tau,\tilde \tau)$ is  determined by the prepotential.  As usual, the \Kahler metric is $\cK_{\ak \tilde \bk} = \p_\ak \p_{\tilde \bk} \cK$;\footnote{We use the notation $\p_\ak \equiv \frac{\p}{\p \tau^\ak}$ and $\p_{\tilde \ak} \equiv \frac{\p}{\p {\tilde \tau}^{\ak}}$.}   
the \Kahler covariant derivative of $X^I$ is $\nabla_\ak X^I = \partial_\ak X^I +\frac12 \cK_\ak X^I$, with $\cK_\alpha = \partial_\alpha \cK$.   

The conjecture may be stated in terms of the boundary limits of the quantities above,\footnote{We use the subscript $_*$ to indicate a quantity evaluated at the boundary, e.g. $X^I_*,\,\, (\nabla_\ak X^I)_* .$} and it employs a complex frame  basis for the boundary \Kahler metric
 \es{FrameDef}{
   e^a_\ak \tilde e^{\tilde b}_{\tilde \bk} \delta_{a \tilde b} = \cK_{\ak \tilde \bk}^* 
 }
with frame fields $e^a_\ak$, their inverses $e^\ak_a$, and their conjugates $\tilde e_{\tilde \bk}^{\tilde b}$.\footnote{In all the examples we consider, we will be able to choose $e^a_\ak$ to be real, so $\tilde e^{\tilde a}_{\tilde \alpha} = e^a_\alpha$.}  The physical real masses of the boundary CFT, called $\mathfrak{m}^a$ in the frame basis, are related to the coefficients of next-to-leading terms in the boundary expansion of $\tau^\ak$
and its conjugate, as explained in more detail in the next section.
The sphere free energy $F_{S^3}(\mathfrak{m})$ can then be immediately obtained from the prepotential ${\cal F}(X^I)$ via  
\es{eq:conjecture}{
	\boxed{F_{S^3}=\frac{2 \pi i L^2}{G_N}{\cal F}(Y^I)\,, \qquad Y^I\equiv X_*^I+\frac{i}{2} e^\ak_a (\nabla_\ak X^I)_*\mathfrak{m}^a \,,}
}
where $G_N$ is the 4d Newton constant. To resolve the projective ambiguity of the $X^I$,  we impose the conditions 
\be
X_*^I= \tilde X_*^I \,, \qquad (\nabla_\alpha X^I)_* = (\nabla_{\tilde \alpha} \tilde X^I)_* \,.\label{eq:Xstar_cond}
\ee

The conjectured relation \eqref{eq:conjecture} provides a far simpler way to find the sphere free energy $F_{S^3}(\mathfrak{m})$ than the traditional method which starts from the BPS equations of a given bulk theory.  Next, one must  obtain the BPS solutions which is never easy, often requiring numerical work. Then come  holographic renormalization of the on-shell action, and implementation of a Legendre transform because the real mass deformation involves some operators whose bulk dual fields obey alternate quantization \cite{Klebanov:1999tb}.  The conjecture delivers the same information simply by substituting $X^I \rightarrow Y^I$ in the prepotential.

\subsection{Background}
\label{sec:background}

That, at leading order in the supergravity approximation, $F_{S^3}(\mathfrak{m})$ is related to the bulk prepotential had also been noticed in certain examples in \cite{Hosseini:2016tor} (see also \cite{Zaffaroni:2019dhb} for a review).  The logic described in \cite{Hosseini:2016tor} is as follows.    First, in explicit 3d ${\cal N} = 2$ SCFTs with holographic duals where the sphere free energy scales as $N^{3/2}$ or $N^{5/3}$ at large $N$, it was noticed that, at leading order in $1/N$, the mass-deformed sphere free energy and the topologically twisted index \cite{Benini:2015noa} on $S^2 \times S^1$ are related.  In particular, the sphere free energy and the ``Bethe prepotential,'' which is an auxiliary quantity used in the index computation, are given by matrix models that can be derived using supersymmetric localization.  At large $N$, these matrix models agree precisely up to normalization.  Second, in supergravity, the topologically twisted index is computed by the entropy of certain extremal black holes.  In the few known top-down examples of such black holes, the attractor mechanism \cite{Ferrara:1996dd} implies that the Bethe prepotential agrees with the prepotential of the supergravity theory.  Thus, in a set of examples where one knows both an explicit Lagrangian description of the SCFT and also the bulk dual of this SCFT in enough detail to construct supersymmetric black hole solutions, Ref.~\cite{Hosseini:2016tor} argued that the sphere free energy is proportional to the bulk prepotential.  One can also incorporate gauged hypermultiplets, at least in principle.  Indeed, in the attractor mechanism, hypermultiplets yield algebraic constraints \cite{Halmagyi:2013sla,Klemm:2016wng}, which suggests that the equivalence between the sphere free energy and the bulk prepotential should hold provided that certain constraints associated with the hypermultiplets are obeyed.

A particularly simple and important example is an ${\cal N} = 2$ supergravity theory with three vector multiplets, no hypermultiplets, and prepotential ${\cal F} = -2i \sqrt{X^0 X^1 X^2 X^3} $.  %
This theory can be embedded as a consistent truncation of ${\cal N} = 8$ gauged supergravity and it has an $AdS_4$ solution of curvature radius $L$.  This theory thus describes a subsector of the $U(N)_k \times U(N)_{-k}$ ABJM theory  \cite{Aharony:2008ug}, when the Chern-Simons level is $k=1$ or $2$ and the ABJM theory has ${\cal N} = 8$ superconformal symmetry.  For ABJM theory  at $k=1$, it was shown in \cite{Jafferis:2011zi} that sphere free energy in this case takes the form $F_{S^3} = \frac{4 \pi N^{3/2}}{3} \sqrt{ 2 \Delta_1 \Delta_2 \Delta_3 \Delta_4}$, where $\Delta_i$ are the R-charges of the bifundamental fields of the theory obeying $\sum_i \Delta_i = 2$.  (As mentioned above, the trial R-charges are related to the real mass deformations by analytic continuation.)  Thus, in this case, it is clear that the sphere free energy is proportional to the prepotential, provided that one identifies the $\Delta_I$ parameters with the $X^I$.  %

Part of our goal in this work is to show that it is possible to prove more directly the relation \eqref{eq:conjecture} between $F_{S^3}(\mathfrak{m})$ and the bulk prepotential.  Indeed, the relation \eqref{eq:conjecture} should follow solely from bulk physics, without invoking either the explicit form of the matrix models obtained after supersymmetric localization in the boundary theory, or the attractor mechanism, or the supersymmetric index computation.  The sphere free energy can be viewed as an observable of the asymptotically AdS bulk theory, so the existence of an SCFT dual should not be necessary.  In the case of ABJM theory, such a direct proof is in fact available:  Ref.~\cite{Freedman:2013oja} constructed asymptotically $AdS_4$ backgrounds in 4d ${\cal N} = 8$ gauged supergravity that are dual to mass deformations of ABJM theory on $S^3$.\footnote{These solutions can be uplifted to solutions of eleven-dimensional supergravity.}  The on-shell action of these solutions does reproduce the sphere free energy $F_{S^3} = \frac{4 \pi N^{3/2}}{3} \sqrt{ 2 \Delta_1 \Delta_2 \Delta_3 \Delta_4}$ given above, and therefore the correspondence between the sphere free energy and the bulk prepotential is explicitly proven without making reference to the properties of the boundary SCFT\@.  We will thus generalize the construction of~\cite{Freedman:2013oja} to other $AdS_4$ supergravity theories and derive the relation \eqref{eq:conjecture} in the examples we study.%

This rest of this paper is organized as follows.  In Section~\ref{sec: abelian vector mults}, we introduce all the ingredients needed to compute the free energy for $AdS_4$ supergravity theories with vector multiplets only, and we check that the direct computation of the free energy agrees with our conjecture in a few examples. In Section~\ref{sec:hypers} we state our conjecture for theories who contain hypermultiplets as well, and work out other examples with one added hypermultiplet. We end with a discussion of our results and future directions in Section~\ref{DISCUSSION}.  Several technical details are relegated to the Appendices.

\section{The free energy in the absence of hypermultiplets}
\label{sec: abelian vector mults}

\subsection{Setting up the computation} \label{sec:sugra_formulae}

Let us begin by considering an $\cN=2$ supergravity theory with $n_V$ Abelian vector multiplets and no hypermultiplets.  The construction of the bulk action for such a theory is well known, so  we give a 
rather minimalist presentation based on Chapter 21 of \cite{Freedman:2012zz}.  With two exceptions, we adhere to the conventions of
this reference:
\begin{enumerate}
	\item The gravitational coupling $1/\kappa^2 = 1/8\pi G_N$ is scaled out as an overall factor of the action. Thus scalar fields are dimensionless.
	\item Starting in Section~\ref{SPHERESLICING}, expressions valid in Lorentzian signature are  converted to Euclidean signature.  Quantities related by complex conjugation in Lorentzian quantum field theory are independent in Euclidean.  We will use $\tilde X^I$ rather than $\bar X^I$ as a  reminder of this fact, and the same for other quantities. We use the conventions for Euclidean supersymmetry from Appendix A of \cite{Freedman:2013oja}.
\end{enumerate}

There are $n_V$ complex scalars $\tau^\alpha$, the ``physical" fields, which parametrize a special K\"{a}hler manifold. The basic data of a specific model is contained in its prepotential ${\cal F}(X^I)$, with $I=0,\ldots,n_V$, a homogeneous function of degree 2 in the $X^I$ which determines the kinetic term and the scalar potential. To relate the $X^I$ to the physical scalars, we use the homogeneous coordinates
\es{XToZ}{
	X^I= y Z^I(\tau^\ak)\,,
}
where the $Z^I(\tau^\ak)$ are holomorphic functions.  

To proceed toward the physical action we need the two matrices:\footnote{Derivatives of the prepotential are indicated by subscripts, e.g. ${\cal F}_I(X)=\frac{\p {\cal F}(X)}{\p X^I}$.} 
\es{NcalN}{
	N_{IJ}&=-i({\cal F}_{IJ}-\bar{{\cal F}}_{IJ})= 2 \Im {\cal F}_{IJ} \,, \\
	\cN_{IJ}(\tau,\btau)&=\bar{{\cal F}}_{IJ}(\btau)+i \frac{N_{IK} Z^K(\tau) N_{JL} Z^L(\tau)}{N_{MN} Z^M(\tau) Z^N(\tau)}\,.
}
Note that ${\cal F}_I=\cN_{IJ}X^J\,.$  The \Kahler potential is then defined by two equivalent formulas:
\es{Kpot}{
	e^{- \cK}=-i (Z^I \bar{{\cal F}}_I(\bZ)-\bar{Z}^I {\cal F}_I(Z))=-N_{IJ}Z^I \bar Z^J \equiv1/y\bar y\,.
}
The last equality defines the product $y \bar y$, leaving the freedom to  redefine  $y \rightarrow t y$, $\tilde y \rightarrow t^{-1}\tilde y$. %
As usual, the Kahler metric is $\cK_{\alpha \bbeta}=\p_\alpha \p_{\bbeta} \cK$.%

In our case of Abelian vector multiplets and no hypermultiplets, the scalar potential is given by
\be
V(\tau ,\btau)=\left( -\frac{1}{2}\left(\Im \cN  \right)^{-1|IJ} -4  X^I \bX^J \right) \vec{P}_I \cdot \vec{P}_J\,. \label{eq:scalar_pot0}
\ee
The $\vec P_I$ are triplet moment maps.  If there are no  hypermultiplets, they are the Fayet-Iliopoulos couplings and must point in  a common direction fixed by the unit vector $\hat e$; hence $\vec P_I = g_I \hat e$ and  $\vec{P}_I \cdot \vec{P}_J = g_I g_J$ \cite{Lauria:2020rhc}. We will make the choice $\hat{e}=(0,0,1)$.

Since the domain wall solutions of our $\cN = 2$ theories involve only scalars and no vectors, it
is valid to truncate to an $\cN = 1$ description for which the 
formalism is far simpler (see Chapter 18 of \cite{Freedman:2012zz}). The $\cN=1$ truncation contains a graviton multiplet (consisting of the metric and a gravitino) and $n_V$ chiral
multiplets (each consisting of a complex scalar and a Weyl fermion).   With the help of (20.190) of \cite{Freedman:2012zz}, the potential can be written in the standard form of an F-term interaction:
\be
V=e^\cK (-3 W \bW+\cK^{\alpha \bbeta} \nabla_\alpha W \nabla_\bbeta \bW)\,,\label{eq:V}
\ee
which contains the usual Kahler covariant derivative:
\beq\label{Kcd}
\nabla_\alpha W=\p_\alpha W+\left( \p_\alpha \cK \right) W\,.
\eeq
Comparing with \eqref{eq:scalar_pot0}, we see that the holomorphic superpotential is related to the $\cN=2$ data by
\be
W(\tau)=g_I Z^I(\tau)\,.  \label{eq:W_Z}
\ee
We can now write the bosonic action of the $\cN=1$ truncation (in mostly plus Lorentzian signature): %
\beq \label{eq:SbulkLor}
S_\text{bulk}=
\int d^4x\,  \sqrt{-g} \, \cL_{\text{bulk}}=
\frac{1}{8 \pi G_N}
\int d^4 x\,  \sqrt{-g}\,  \left[\frac12\cR -\cK_{\alpha \bar{\beta}} \p_\mu \tau^{\alpha} \p^\mu \bar \tau^{\bar{\beta}} -V(\tau,\bar \tau) \right] \,.
\eeq

\subsection{$AdS_4$ solution, ${\cal N} = 2$ SCFT interpretation, and alternate quantization}
\label{ALTERNATE}

In the examples we study, the potential $V(\tau, \bar \tau)$ has a critical point at $\tau^\ak = \tau^\ak_*,~ \bar \tau^{\bar\bk}= \bar \tau^{\bar \bk}_*$, which yields an $AdS_4$ solution of the theory \eqref{eq:SbulkLor}, where
\es{AdSSoln}{
	g_{\mu\nu} = g_{\mu\nu} \big|_\text{AdS} \,, \qquad
	\tau^\ak = \tau^\ak_* \,, \qquad  \bar \tau^{\bar\bk}= \bar \tau^{\bar \bk}_* \,,
}
and all other fields vanish. This supersymmetric  critical point is defined by the $2n_V$ conditions $\nabla_\ak W=\nabla_{\bar\bk} \bar W=0$; these, in turn, imply that $\p_{\alpha}V=\p_{\bar \beta}V=0$. The value of the potential at the critical point is then related to the AdS curvature radius $L$  by 
\be  \label{V*L}
V_*= \frac{-3}{L^2}  = -3e^{\cK_*} W(\tk_*)\bar W(\bar \tk_*)\,.
\ee
In the ${\cal N} = 2$ theory that truncates to \eqref{eq:SbulkLor}, the condition \eqref{V*L} reads $\abs{g_I X^I_*}^2 = 1/L^2$, as can be seen from combining \eqref{V*L} with~\eqref{XToZ}, \eqref{Kpot}, and \eqref{eq:W_Z}.  Here and in the following discussion, we denote with a subscript ${}_*$ quantities evaluated at the critical point of the potential, namely  $X^I_* = X^I(\tau_*, \bar \tau_*)$, etc.  Upon using the transformation $y \rightarrow t y$, $\tilde y \rightarrow t^{-1}\tilde y$ with $\abs{t} = 1$, we can assume without loss of generality that
\es{Choice}{
	g_I X^I_* = g_I \bar X^I_* =  \frac{1}{L} \,.
}
In the ${\cal N} = 2$ theory, this AdS solution preserves eight linearly-independent supersymmetries.  With the choice in \eqref{Choice}, the Killing spinor equations obtained from the vanishing of the gravitino variations $\delta \psi_\mu^i = \delta \psi_{i \mu}  = 0$ (see Eq.~(21.42) of \cite{Freedman:2012zz}) take the form
\es{KSForm}{
	D_\mu \epsilon^i = - \frac{1}{2L} \gamma_\mu \tau_3^{ij} \epsilon_j \,, \qquad
	D_\mu \epsilon_i = - \frac{1}{2L} \gamma_\mu \tau_{3 ij} \epsilon^j \,,
}
where $i, j$ are $SU(2)_R$ indices raised and lowered\footnote{For details, see Appendix 20A of \cite{Freedman:2012zz}.} with the epsilon symbol, and $(\tau_3)_i{}^j \equiv i (\sigma_3)_i{}^j$.

Via the AdS/CFT duality, the $AdS_4$ solution is dual to a 3d ${\cal N} = 2$ SCFT that lives on the boundary in cases where the model \eqref{eq:Sbulk} can be obtained from a top-down string theory or M-theory construction.  We do not attempt to embed the model \eqref{eq:Sbulk} into ten-dimensional or eleven-dimensional supergravity, however, so the existence of a boundary SCFT is an assumption that we make.   The fluctuations of the metric, scalar fields, gauge fields, gravitino, and fermions are dual to certain operators in the boundary SCFT\@.  In particular, these operators belong to the stress tensor multiplet of the boundary SCFT, which is dual to the ${\cal N} = 2$ Weyl multiplet in the bulk, as well as $n_V$ conserved current multiplets, which are dual to the $n_V$ vector multiplets.

An important subtlety in the details of the AdS/CFT dictionary involves the boundary conditions on the scalar and fermion fields.  The $n_V$ vector multiplets contain $2n_V$ real scalars and $2n_V$ Majorana fermions.  When expanded around the AdS solution to quadratic order, one can prove that all the scalar fields are conformally coupled, i.e.~they have squared mass $m^2 = -2/L^2$, and all the fermions are massless---see Appendix~\ref{app:mass spectra}\@.  In AdS, scalar fields of this mass can obey regular boundary conditions or alternate boundary conditions depending on whether they are dual to boundary operators of dimension $2$ or $1$, respectively \cite{Klebanov:1999tb}.  In fact, both cases should occur for us, because each of the $n_V$ conserved current multiplets in the boundary theory contains a scalar superconformal primary $J^\alpha$ of dimension $1$ and another scalar conformal primary $K^\alpha$ of dimension $2$.  The question then is:  which scalar fields in the bulk obey regular boundary conditions, and which ones obey alternate boundary conditions?  A similar question can be asked about the fermions.   The massless bulk fermions correspond to dimension $3/2$ operators in the boundary theory.  In the bulk it is a priori not clear which boundary components of the fermions correspond to field theory sources and which ones correspond to VEVs.  It is important that no similar ambiguity occurs for the gauge fields, gravitino, or the metric fluctuations, which all obey the regular quantization that identifies the coefficient of the leading behavior close to the boundary as the source for the dual boundary operator.

The boundary conditions for scalars and fermions can be determined by examining the asymptotic behavior of the fluctuations around the $AdS_4$ solution and how the various coefficients in the asymptotic expansion transform under supersymmetry.  The guiding principle is that, under supersymmetry, sources for boundary operators should transform into sources.  Since we know which coefficients in the asymptotic expansions of the guage fields correspond to sources for the boundary conserved currents, the supersymmetry transformations would then determine which coefficients should be interpreted as sources for the fermions and the scalar fields as well.  Note that ${\cal N} = 2$ supersymmetry is crucial in this case, because ${\cal N} = 1$ supersymmetry transformations do not relate scalars to gauge fields.

We perform the analysis outlined above in Appendix~\ref{ASYMPTOTIC}\@.  For the scalar fields, the result is that 
\es{Regular}{
	{\cal B}^I \equiv \Im \left( ( \tau^\alpha - \tau^\alpha_*) (\nabla_\alpha X^I)_* \right) 
}
should be quantized with regular quantization, while
\es{Alternate}{
	{\cal A}^I \equiv \Re \left( ( \tau^\alpha - \tau^\alpha_*) (\nabla_\alpha X^I)_* \right) 
} 
should obey alternate quantization.  Alternatively, one can say that $\Im ( X^I - X^I_*)$ obeys regular boundary conditions while $\Re (X^I - X^I_*)$ obeys alternate boundary conditions.

In practice and with no loss of generality, one can take $ \tau^\alpha_* = 0$ and $(\nabla_\alpha X^I)_*$ real, and then $\Im \tau^\alpha$ will be regularly quantized while $\Re \tau^\alpha$ will obey alternate quantization.  In this simplified setup, the correspondence between bulk fields and SCFT operators is given in Table~\ref{tab:dictionary}.  We will discuss the precise normalization of the boundary operators $J^\alpha$ and $K^\alpha$ later on in Section~\ref{REALMASSDEF}.
\begin{table}[ht]
	\begin{center}
		\begin{tabular}{c|c|c}
			operator & dual field & $\Delta$ \\
			\hline
			$J^\alpha$ & $\Re \tau^\alpha$ & $1$  \\
			$K^\alpha$ & $\Im \tau^\alpha$ & $2$  
		\end{tabular}
	\end{center}
	\caption{Correspondence between bulk fields and boundary operators in the ${\cal N} = 2$ SCFT\@.}
	\label{tab:dictionary}
\end{table}%

\subsection{Solutions with sphere slicing}
\label{SPHERESLICING}

In the rest of this section we work in Euclidean signature, where the action~\eqref{eq:SbulkLor} becomes
\beq \label{eq:Sbulk}
S_\text{bulk}=\frac{1}{8 \pi G_N}
\int d^4 x \sqrt{g} \left[-\frac12\cR +\cK_{\alpha \tilde{\beta}} \p_\mu \tau^{\alpha} \p^\mu \tilde \tau^{\tilde{\beta}} +V(\tau,\tilde \tau) \right] \,.
\eeq
As mentioned in the Introduction, we are interested in constructing classical solutions of the theory \eqref{eq:Sbulk} that correspond to real mass deformations of the putative dual ${\cal N} = 2$ SCFT on $S^3$.  Such solutions have an $SO(4)$-invariant asymptotically AdS metric tensor that conforms to the ansatz (see \cite{Freedman:2013oja}):
\be
ds^2=L^2 e^{2A(r)} ds^2_{S^3}+e^{2B(r)}dr^2\,. \label{eq:metric_ansatz}
\ee
The frame fields are
\be
e^i = L e^{A(r)}\hat{e}^i \qquad e^4 =e^{B(r)}dr\,, \label{eq:frame}
\ee
where the index $i$ runs over $i=1,2,3$ and the $\hat{e}^i $ are a choice of $S^3$ frame fields. With $\hat \omega^{ij}$ denoting an $S^3$ connection, the spin connection for \eqref{eq:frame} is %
\be\label{eq:conn}
\omega^{ij}=\hat\omega^{ij},  \qquad \omega^{i4}=A' e^{-B}e^i\,.
\ee
It is redundant to specify two radial functions in \eqref{eq:metric_ansatz},   but it is convenient since we will use two different gauges. The conformally flat (CF) gauge, with $e^{B(r)}=\frac{L}{r}e^{A(r)}$, is somewhat more convenient in the search for analytic solutions of the BPS equations, while
the Fefferman-Graham (FG) gauge, with $B=\log L$,  is better suited for numerics and for holographic renormalization.

The scalar fields $\tau(r),~\ttk(r)$ of the domain wall approach a supersymmetric critical point of the potential \eqref{eq:V} at the AdS boundary, i.e.~$\tau^\ak(r) \rightarrow \tau^\ak_*,~ \ttk^{\tilde\bk}(r)\rightarrow \ttk^{\tilde\bk}_*$  as $r \rightarrow \infty$  in the FG gauge.  The real mass parameters appear in the asymptotic expansion close to the boundary of the fields $\tau^\alpha$ and $\tilde \tau^\alpha$, as made explicit below.

\subsubsection{The BPS equations}

The next item of business is the BPS equations satisfied by supersymmetric $SO(4)$-invariant solutions of the theory.  These are first order partial differential equations  derived from the requirement that SUSY variations of the fermions vanish.  Any solution of the BPS equations is also a solution of the bosonic equations of motion.  Our discussion follows the treatment in \cite{Freedman:2013oja}.  In the Euclidean signature $\cN=1$ truncation of a theory without hypermultiplets, the BPS equations are
\beg
\delta \psi_\mu = \left( \p_\mu + \frac{1}{4} \omega_{\mu}^{ab}\sigma_{[a} \bar{\sigma}_{b]}- 
\frac{i}{2} \cA_\mu \right) \epsilon + \frac{1}{2}  \sigma_\mu e^{\cK/2} W \tilde{\epsilon}=0 \,, 
\\
\delta \tilde \psi_\mu= \left( \p_\mu + \frac{1}{4} \omega_{\mu}^{ab} \bar\sigma_{[a} {\sigma}_{b]}+ 
\frac{i}{2}\cA_\mu \right) \tilde \epsilon + \frac{1}{2}\bar\sigma_\mu e^{\cK/2} \tilde W {\epsilon} =0 \,, \\
\delta \chi^\alpha = \sigma^\mu \p_\mu \tau^\alpha \tilde{\epsilon}-e^{\cK/2}g^{\alpha { \beta}}\tilde \nabla_{ \beta} \tilde W \epsilon=0 \,, \\
\delta \tilde \chi^\beta = \bar \sigma^\mu \p_\mu \tilde \tau^\beta {\epsilon}-e^{\cK/2}g^{\alpha \beta}\nabla_\alpha  W \tilde \epsilon 
=0
\label{eq:susy_var}\,,
\eeg
where $\cA_\mu$ is the \Kahler connection, $\cA_\mu = \frac{i}{2}\left(\p_\mu \tau^\alpha \p_\alpha \cK-\p_\mu \tilde \tau^\alpha \tilde{\p}_\alpha \cK \right)$.

We are interested in solutions where the scalars $\tk^\ak(r)$ are strictly radial and the frame and connection come from \eqref{eq:metric_ansatz}--\eqref{eq:conn} above.  The 2-component spinors $\epsilon$ and $\tilde\epsilon$ are Killing
spinors.  Due to rotational symmetry,  a given solution has either the structure $\epsilon = i(r)\zeta,~ \tilde\epsilon = \tilde i(r)\zeta$ \,or\, 
$\epsilon = j(r)\xi,~ \tilde\epsilon = \tilde j(r)\xi$,  where $\zeta,~ \xi$ are Killing spinors of $S^3$ and satisfy 
\be\label{eq:sphere_spinors}
\nabla_i \zeta= \frac{i}{2} \sigma_i \zeta\,, \qquad \quad \nabla_i \xi=- \frac{i}{2} \sigma_i \xi\,.
\ee
The choice of the $\zeta$ or $\xi$ structures determines whether the solution is invariant under the left or right $SU(2)$ factor  of the  isometry group of the sphere, $SO(4)=SU(2)_l \times SU(2)_r$. We choose the $\zeta$ structure and look for solutions of the BPS equations in this case. (In solutions with the $\xi$ structure, the roles of $\tau^\alpha$ and $\tilde \tau^\alpha$ are switched.)

We now group the BPS equations for the spin 1/2 fermions and the gravitinos (for $\mu=i$ 1,2, or 3) into a $(2n_V+2)\times 2$ dimensional matrix. After insertion of $A(r),\, B(r)$ from \eqref{eq:metric_ansatz} and \eqref{eq:frame} and use of \eqref{eq:sphere_spinors}, we obtain
\be
\begin{pmatrix}
	1+L A' e^{A-B} & -i L e^A W e^{\cK/2}\\
	-i  L e^A \tilde{W} e^{ \cK/2} & 1-L A' e^{A-B} \\
	-e^{\mathcal{K}/2} \cK^{\alpha \beta} \tilde \nabla_{ \beta} \tilde{W}& -i e^{-B} (\tau^\alpha)'\\
	i e^{-B} (\tilde{\tau}^\beta)' & -e^{\mathcal{K}/2} \cK^{\alpha \beta}   \nabla_{ \alpha} {W}
\end{pmatrix}
\begin{pmatrix}
	\epsilon\\
	\tilde{\epsilon}
\end{pmatrix}
=0 \,. \label{eq:BPS_gen}
\ee

A non-trivial solution of these equations is possible only if all $2 \times 2$ minors of this matrix vanish. In this way we look for solutions for the scalar fields $\tau^\alpha$, $\tilde{\tau}^\alpha$ and the metric functions. Once this is achieved, we can rewrite the BPS equations  $\delta \psi_r=0$ and $\delta \tilde \psi_r =0$ as the $2\times 2$ linear system that determines the spinors $\epsilon$ and $\tilde \epsilon$
\beq
\p_r \begin{pmatrix}
	\epsilon\\
	\tilde{\epsilon}
\end{pmatrix} =
\begin{pmatrix}
	\frac{1}{2}i \cA_r & \frac{i}{2}  e^B W e^{ \cK/2}\\
	- \frac{i}{2} e^B \tilde{W} e^{\cK/2} &-\frac{1}{2}i \cA_r
\end{pmatrix}
\begin{pmatrix}
	\epsilon\\
	\tilde{\epsilon}
\end{pmatrix} \,.
\label{eq:BPS_r} 
\eeq

\subsubsection{Computation of the renormalized on-shell action:}
\label{sec:free_energy}
The next step in the evaluation of the sphere free energy is to compute the on-shell action obtained by substitution of a solution of the field equations, in our case a solution of the BPS equations,  in the bulk action $S_{\rm bulk}$ of \eqref{eq:Sbulk}.  However, the radial integral diverges near the AdS boundary.  So the integral must be cut off at $r=r_c$ and boundary counterterms added to cancel the divergences.  This is the well known process of holographic renormalization \cite{Skenderis:2002wp}, adapted to our application in Section~6.1 of
\cite{Freedman:2013oja}.

As usual when working in a spacetime with boundary, we introduce the Gibbons-Hawking-York term
\be
S_{\text {GHY}}=-\frac{1}{8\pi G_N} \int_{\p} d^3 x\left. \sqrt{h} \, K \right|_{r=r_c} \,,
\ee
where $h_{ij}$ is the induced metric at the cutoff and $K$ is the trace of the extrinsic curvature.\footnote{In the FG gauge, where the metric behaves asymptotically as $ds^2 = L^2\left( dr^2+\frac{e^{2r}}{4}d \Omega_{S^3}^2\right)$, then the induced metric is $h_{ij} =(Le^{r_c}/2)^2h_{0ij}$, where $h_{0ij}$ is a metric on the unit sphere. Then $K=L^{-1}\p_{r_c} \log \sqrt{h}$.}
Its role is to provide a well defined variational principle for the $\sqrt{g}R$ term in the bulk Lagrangian.\,%
We then need two other counterterms, the first because the boundary metric is curved with Ricci scalar $\cR_h =24 e^{-2r_c}/L^2$:
\beg
S_{h}=\frac{L}{16\pi G_N} \int_{\p} d^3 x\, \left.\sqrt{h}\, \cR_h \right|_{r=r_c}\,.  \label{eq:Sh}
\eeg
The second is needed to cancel divergences while maintaining supersymmetry:
\beg
S_{\text{SUSY}}= \int_\p d^3 x \, \sqrt{h} \, \cL_{\text{SUSY}} = \frac{1}{4\pi G_N}\int_{\p} d^3 x\, \left. \sqrt{h}\, e^{\cK/2} |W| \right|_{r=r_c}  \,.\label{eq:Ssusy}
\eeg

Let us begin to put these ingredients together, working with the redundant metric \eqref{eq:metric_ansatz}. The goal is to obtain an integral formula  for the sum of the on-shell action  plus the three counterterms which converges at short and long distances and depends only on the metric functions $A(r), \, B(r)$.  Since fields and metric depend only on $r$, the integration over coordinates of $S^3$ produces the volume factor of $h_{0ij}$ which is $2\pi^2$.  The first step is to take $S_{\text{bulk}} + S_{\text{GHY}} $ and integrate by parts to cancel the $A''$ term. The boundary term is cancelled by the counterterm, leaving
\beq
S_{\text{bulk}}+S_{\text{GHY}}= \frac{ \pi L}{4G_N}  
\int_{r_{\text{IR}}}^{r_c} dr e^{A+B} \left[-3\left(1+(e^{A-B}L A')^2 \right)  
+L^2 e^{2A} \left(e^{-2B} \cK_{\alpha \beta}\p_r \tau^\alpha \p_r \tilde\tau^\beta+V  \right)\right] \,.
\label{eq:bulk+ghy}
\eeq
The Euler variation of this expression with respect to $A(r)$ then produces 
\beq
L^2 e^{2A+2B} (e^{-2B} \cK_{\alpha \beta} \p_r \tau^\alpha \p_r \tilde\tau^\beta+V)=e^{2B}-L^2 e^{2A}\left(3 (A')^2-2 A' B'+2 A'' \right) \,.
\eeq
We insert this in the integrand of \eqref{eq:bulk+ghy}, which then reads
\beq
\left. S_{\text{bulk}}+S_{\text{GHY}}\right|_{\text{on-shell}}=-\frac{\pi L}{2G_N} \int_{r_{\text{IR}}}^{r_c}  d r\, e^{A+B} \left[1+L^2 e^{2A-2B}(3 (A')^2-A' B'+A'')  \right]  \,. \label{eq:SA_1}
\eeq

Now let's look at the counterterms. We can simplify $S_{\text{SUSY}}$ using the BPS equations. In particular the vanishing of the determinant of the first two lines of \eqref{eq:BPS_gen} gives the equation
\beq
1-(L A' e^{A-B})^2+ L^2 e^{2A}e^{ \cK} W \tilde W=0\,. \label{eq:A_W_relation}
\eeq
Thus we can write
\beq
S_{\text{SUSY}}=\frac{1}{4\pi G_N} \int_{S^3} d^3 x\, \left. \sqrt{h}\, e^{ \cK/2} |W| \right|_{r=r_c}=\frac{\pi L^2}{2G_N}\left[e^{2A} \sqrt{(L A' e^{A-B})^2 -1}\right]_{r=r_c}\,.
\eeq
Similarly,    %
\be
S_h=\frac{\pi L^2}{8 G_N}3e^{A(r_c)}\,.
\ee
The sum of $S_{\text{bulk}}$ plus all counterterms can then be written as
\ie
S_{\text{reg}}=-\frac{\pi L}{2G_N} \lim_{r_c \to r_{\text{UV}}}&\left(   \int_{r_{\text{IR}}}^{r_c} d r\, e^{A+B} \left[1+L^2 e^{2A-2B}(3 (A')^2-A' B'+A'') \right]\right.-
\\
& \qquad \left.-L e^{3A}\left[\frac{3}{2} e^{-2A}+\sqrt{(L A' e^{-B})^2 -e^{-2A}}\right]_{r=r_c}\right) \,. \label{eq:Sreg}
\fe

Further simplification occurs in the FG gauge, where $B=\log L$.  We can use the asymptotic form of $A$ for large $r$,
\be
e^{2A}=\frac{e^{2r}}{4}+\text{constant}+\ldots\,,
\ee
to further simplify the boundary term
\beq
e^{3A}\left[ \frac{3}{2} e^{-2A}+\sqrt{(A')^2-e^{-2A}} \right] \sim e^{3A} A'+ e^A +O(e^{-r})\,.
\eeq
Then \eqref{eq:Sreg} becomes
\beq
S_{\text{reg}}=-\frac{\pi L^2}{2G_N} \lim_{r_c\to \infty}\left[\int_0^{r_c} dr\,\left(  e^A \left(1+3 e^{2A} (A')^2+e ^{2A} A'' \right)\right) - e^A\left.\left( e^{2A}  A'+ 1\right) \right|_{r_c}\right]
\eeq
We can write the boundary term as an integral of a total derivative, with no contribution in the IR, since $e^{A(r)} \sim r$ as $r\rightarrow 0 $ is required for a non-singular bulk metric. Sweet cancellations occur and we are left with the convergent final result for the renormalized on-shell action:
\beq
S_{\text{reg}}=\frac{ \pi L^2}{ 2G_N} \int_0^\infty dr\, e^{A}(A'-1)\,. \label{eq:Sreg_FG}
\eeq
\subsubsection{Legendre transform}
\label{LEGENDRE}

It might be tempting to identify $S_{\text{reg}}$ with the sphere free energy, but one needs to be careful because, as discussed in Section~\ref{ALTERNATE}, some of the bulk scalar fields are dual to boundary operators of dimension $1$ and therefore require alternate quantization.  Working under the simplifying assumptions $ \tau^\alpha_* = 0$ and $(\nabla_\alpha X^I)_*$ real mentioned at the end of Section~\ref{ALTERNATE}, the fields requiring alternate quantization are $\tau^\alpha + \tilde \tau^\alpha$;  these fields are the Euclidean continuations of $2\Re \tau^\alpha$ from Lorentzian signature.  As instructed by \cite{Klebanov:1999tb}, the free energy in the boundary theory is given by the Legendre transform of $S_{\text{reg}}$ with respect to the leading coefficients in the asymptotic expansions of $\tau^\alpha + \tilde \tau^\alpha$.

In more detail, a general $SO(4)$-invariant solution of the second order equations of motion in the FG gauge has, close to the boundary, the following asymptotic form
 \es{eq:FG_asymptotics}{
\tau^\alpha = a^\alpha e^{-r}+b^\alpha e^{-2r}+\ldots \,, \\
\tilde \tau^\alpha = \tilde a^\alpha e^{-r}+\tilde b^\alpha e^{-2r}+\ldots \,.
 }
The on-shell action $S_\text{reg}$ is naturally a function of $(a^\alpha, \tilde a^\alpha)$ (as well as other boundary sources) because when deriving the second order equations of motion from the Lagrangian, one has to hold $(a^\alpha, \tilde a^\alpha)$ fixed.  The alternate quantization procedure amounts to a Legendre transformation that takes us from $S_\text{reg}(a^\alpha, \tilde a^\alpha)$ to a function of $a^\alpha - \tilde a^\alpha$ and the canonically conjugate variable to $a^\alpha + \tilde a^\alpha$:
\be
F_{S^3}=S_{\text{reg}}(a^\alpha, \tilde a^\alpha)-\frac{1}{2}\sum_\alpha  (a^\alpha+\tilde{a}^\alpha) \left(\frac{\partial S_{\text{reg}}}{\partial {a}^\alpha}+ \frac{\partial S_{\text{reg}}}{\partial \tilde{a}^\alpha}\right)\,. \label{eq:F_Legendre}
\ee
This is our final formula that we will evaluate in  several examples below.  To compute the variations of $S_{\text{reg}}$ with respect to the $a^\alpha$'s, it is convenient to use the fact that
\be
\frac{\partial S_\text{reg}}{\partial a^\alpha}=2 \pi^2 L^3 \lim_{r \to \infty}e^{-r} e^{3A}\left(L \frac{\p \mathcal{L}_{\text{bulk}}}{\p (\p_r \tau^\alpha)}+\frac{\p \cL_{\text{SUSY}}}{\p \tau^\alpha}\right)\,, \label{eq:Leg_var}
\ee
which will be written out explicitly on a case by case basis.  

\subsubsection{Real mass deformation in the boundary theory}
\label{REALMASSDEF}

Let us now make the connection between the asymptotic expansion \eqref{eq:FG_asymptotics} and the boundary real mass parameters $\mathfrak{m}^a$ appearing in our conjecture \eqref{eq:conjecture}.  As already mentioned, our work does not necessarily assume that the bulk supergravity theory is dual to a boundary QFT, so the boundary mass parameters $\mathfrak{m}^a$ written below can simply be taken as a definition.  However, in cases where there is a boundary dual description, the definition of $\mathfrak{m}^a$ is motivated by the field theory description, so let us take that perspective in the following discussion.

As discussed in Section~\ref{ALTERNATE}, each vector multiplet in the bulk corresponds to a conserved current multiplet, which contains scalar operators of dimension $1$ and $2$.  Let us denote these scalar operators by $J^a$ and $K^a$ and normalize them so that, in flat space, they give the two-point functions %
 \es{normalization}{
   \langle J^a(\vec{x}) J^b(0) \rangle &= \frac{{\cal C} \delta^{ab}}{16 \pi^2 \abs{\vec{x}}^2} \,, \qquad
    {\cal C} \equiv \frac{L^2}{2 \pi G_N} \\
   \langle K^a(\vec{x}) K^b(0) \rangle &= \frac{{\cal C} \delta^{ab}}{8 \pi^2 \abs{\vec{x}}^4} \,.
 }
With this normalization, after mapping these operators to an $S^3$ of unit radius, we can write the real mass deformation as \cite{Closset:2012vg}
 \es{RealMass}{
   \sum_a  \mathfrak{m}^a \int_{S^3} d^3\vec{x}\, \sqrt{g(\vec{x})} \left[ i J^a(\vec{x}) + K^a(\vec{x}) \right]  \,.
 }
In particular, the mass parameters $\mathfrak{m}^a$ are sources for the dimension $2$ operators $K^a$.  Thus, they must be linear combinations of the coefficients of $e^{-r}$ in the boundary expansion of $\tau^\alpha - \tilde \tau^\alpha$, since it is the $\tau^\alpha - \tilde \tau^\alpha$ fields that are regularly quantized.

In order to relate the boundary mass parameters to the asymptotic expansions \eqref{eq:FG_asymptotics}, we first rescale the scalar fields $\tau^\alpha$ and $\tilde \tau^\alpha$ so that, close to the boundary, they become canonically normalized.  This is achieved by defining 
 \es{tauaDef}{
  \tau^a \equiv e^a_\alpha \tau^\alpha \,, \qquad
   \tilde \tau^a \equiv e^a_\alpha \tilde \tau^\alpha \,,
 }
where the frame fields are defined in \eqref{FrameDef}, and we assumed that $e^a_\alpha$ is real, as will be the case in all our examples.  Because $\tau^\alpha - \tilde \tau^\alpha$ is regularly quantized then so is $\tau^a - \tilde \tau^a$.  This means that the coefficient of $e^{-r}$ in the large $r$ expansion of $\tau^a - \tilde \tau^a$ is proportional to the source for the dual operator $K^a = e^a_\alpha K^\alpha$, which is the mass parameter $\mathfrak{m}^a$.  In particular, we define
 \es{Gotma}{
  \mathfrak{m}^a = e^a_\alpha \frac{a^\alpha - \tilde a^\alpha}{2 i} \,.
 }
Here, the normalization was chosen in a way consistent with \eqref{normalization}--\eqref{RealMass}.  In particular, from \eqref{RealMass}, it follows that to quadratic order in the small $\mathfrak{m}^a$ expansion, we have\footnote{We take the real part of $F_{S^3}$ because, as explained in \cite{Closset:2012vg}, the imaginary part may also receive scheme-dependent contributions from contact terms in the $\langle J^a K^b \rangle$ correlators.}
 \es{derFS3}{
  \frac{\partial^2 \Re F_{S^3}}{\partial \mathfrak{m}^a \partial \mathfrak{m}^b}
   =  \int_{S^3} d^3\vec{x}\, \sqrt{g(\vec{x})}  \int_{S^3} d^3\vec{y}\, \sqrt{g(\vec{y})}
    \left[ \langle J^a(\vec{x}) J^b(\vec{y}) \rangle -  \langle K^a(\vec{x}) K^b(\vec{y}) \rangle\right] \,.
 } 
With \eqref{normalization}, the integrals over $S^3$ can be performed as in  \cite{Closset:2012vg} with the result
 \es{derFS3Again}{
  \frac{\partial^2 \Re F_{S^3}}{\partial \mathfrak{m}^a \partial \mathfrak{m}^b}
   = \delta^{ab} \frac{\pi^2}{2} {\cal C} = \delta^{ab} \frac{L^2 \pi}{4 G_N} \,.
 }
We will see that the normalization in \eqref{Gotma} ensures that \eqref{derFS3Again} is obeyed in all models we study.

With this definition, the sphere free energy $F_{S^3}$ from \eqref{eq:F_Legendre} can be expressed in terms of $\mathfrak{m}^a$.  As we will see in the examples below, the frame vectors $e^a_\alpha$ will cancel in all practical computations, and the free energy will take a simpler form in terms of $\mathfrak{m}^a$.

\subsubsection{Independence of the FI parameters} \label{sec:FI_independence}

When written in terms of $\mathfrak{m}^a$ and $L$, the free energy $F_{S^3}$ should not depend on the FI parameters $g_I$.  As discussed in the Introduction, this follows from the general arguments presented recently in \cite{Binder:2021euo}, and we will check it for generic $g_I$ in examples in Sections~\ref{sec:eta_vectors} and~\ref{sec:STU} and Appendix~\ref{app:general_couplings}.  One practical consequence is that if we are interested in the expression for $F_{S^3}$ in a given supergravity theory, we do not have to compute it for arbitrary $g_I$. Instead, we can compute it for a convenient choice of $g_I$ that simplifies the computation. %

\subsection{$\cF=\frac{i}{4} \eta_{IJ} X^I X^J$} \label{sec:eta_vectors}
The first example we study is a model with $n_V$ multiplets, described by the quadratic prepotential
\be\label{Fquad}
\cF=\frac{i}{4} \eta_{IJ} X^I X^J\,,
\ee
where $I=0,\ldots,n_V$ and $\eta_{IJ} = \text{diag}(-,+,\ldots,+)$. Black hole solutions in this model were studied in \cite{Cacciatori:2009iz}.
It is convenient to parametrize the $Z^I$ coordinates as
\be
Z^I=\left(1-\frac{1}{g_0}\vec{g}\cdot \vec{\tau}, \tau^i-\frac{g_i}{g_0} \right) \,,  \label{eq:eta_Z_param}
\ee
where $i=1,\ldots,n_V$ and the scalar product denotes $\vec{g}\cdot \vec{\tau}=g_i \tau^i$. This parametrization was chosen after some exploration;  one advantage is that the fields $\tau^i$ and $\tilde{\tau}^i$ vanish at the AdS boundary (as we will see).

Using the formulae in Section~\ref{sec:sugra_formulae}, we find that
\be
\cF_I(Z)=\frac{i}{2}Z_I \,,
\ee
where the $I$ index of $Z^I$ is lowered by $\eta_{IJ}$, and the \Kahler potential is
\be
 e^{- \cK}=-Z^I \bZ_I=\frac{g^2+|\vec{g}\cdot \vec{\tau}|^2  -g_0^2|\vec{ \tau}|^2}{g_0^2} \,,
\ee
with  $g^2\equiv -g_I g_J \eta^{IJ}$. Another advantage of the choice \eqref{eq:eta_Z_param} is that the superpotential is particularly simple,
\beq
W=g_I Z^I=\frac{g^2}{g_0}\,.
\eeq
The scalar potential, computed using \eqref{eq:V}, is
\be
V=-g^2 \frac{3 g^2+|\vec{g}\cdot \vec{\tau}|^2-g_0^2 |\vec{\tau}|^2}{g^2+|\vec{g}\cdot \vec{\tau}|^2-g_0^2 |\vec{\tau}|^2}
=  -g^2\left[3 + 2\frac{g_0^2 |\tau|^2 -|\vec g\cdot\tau|^2}{g^2 -g_0^2 |\tau|^2 +|\vec g\cdot\tau|^2}\right] \,.
\ee
The unique extremum\footnote{This extremum is a maximum, and the mass eigenvalues are $m^2=-2/L^2$, as expected for the scalars of vector multiplets.}  of the potential is at $\tau^i=0$, where it takes the value (see \eqref{V*L}):
\be
V_*=-3 g^2 =-3/L^2\,.
\ee
This determines the radius of AdS as a function of the couplings, and we require $g^2>0$, as appropriate for AdS.

We now have all the ingredients needed to search for non-trivial solutions of the BPS equations. We consider the $2\times 2$ minors of \eqref{eq:BPS_gen} as well as the equations of motion obtained by extremizing \eqref{eq:Sbulk}. The first step is to set up a series expansion around $r=0$, and solve for the first few orders.  We require regularity at the origin;  the scalars $\tau^i$ are finite at $r=0$ while $e^A(r) \sim r$ in the CF coordinates that we use.  
In this model the series expansion allows us to  guess the full solutions.%

The non-trivial solution to the BPS equations \eqref{eq:BPS_gen} and \eqref{eq:BPS_r} is 
\be
\tau^\alpha= c^\alpha(1-r^2) \,, \qquad \tilde{\tau}^\alpha=0\,,  \qquad e^{2A}=  \frac{4r^2}{(1-r^2)^2}  \,, \label{eq:eta_sols}
\ee
with $c^\ak$ arbitrary constants,
and the Killing spinors are %
\be
\begin{pmatrix}
	\epsilon\\
	\tilde{\epsilon}
\end{pmatrix}= \frac{1}{\sqrt{1-r^2}} 
\begin{pmatrix}
	r\\
	-i
\end{pmatrix} \zeta\,.
\ee
If we were to consider the Killing spinors  proportional to $\xi$ (see \eqref{eq:sphere_spinors}), then we would find a solution with the $\tau^\alpha$ and $\tilde{\tau}^\alpha$ swapped.

The metric tensor of this solution is undeformed Euclidean AdS\@.  There is  no back-reaction on the metric, despite the presence of non-constant scalar fields. This might seem surprising at first, but it is allowed in Euclidean signature because
the fields $\tau^\alpha$ and $\ttk^\alpha$ are independent.  The scalar stress tensor contains only products such as $\tau^\alpha \tilde{\tau}^\beta$ or $\partial_\mu \tau^\alpha \partial_\nu \tilde \tau^\beta$. Since the $\ttk$ fields vanish,  there is no back-reaction on the metric.  This behavior would not be possible in Lorentzian signature where $\tilde \tau^\alpha$ and $\tau^\alpha$ are related by complex conjugation.

We now compute the free energy of the boundary theory. For this we pass to the FG gauge, via the  coordinate transformation is $r \to \tanh r /2$, and evaluate  \eqref{eq:Sreg_FG}.    The transformed metric is
\be
ds^2
=L^2\left( dr^2+e^{2A(r)}\; ds^2_{S^3}\right) \,.
\ee
and the BPS solutions are
\be
\tau^\alpha =c^\alpha \sech^2 \frac{r}{2}\,, \qquad \tilde \tau^\alpha=0\,, \qquad e^{2A}=\sinh^2r\,. \label{eq:bps_eta}
\ee
We get
\be
S_{\text{reg}}= %
\frac{\pi L^2 }{2G_N}\,. \label{eq:S_onshell}
\ee

The next step is the Legendre transform.   To prepare for this we record the asymptotic form of the scalars and identify $a^{\alpha}$ and $b^\alpha$:
\be
\tau^\alpha =4 c^\alpha e^{-r}-8 c^\alpha e^{-2r}+\ldots\equiv a^\ak e^{-r} + b^{\ak}e^{-2r}\,. \label{eq:eta_asymp}
\ee
Since the $\ttk^\ak$ vanish, the general form in \eqref{eq:F_Legendre} reduces to
\be
F_{S^3}=S_{\text{reg}}-\frac{1}{2}\sum_\alpha a^\alpha \frac{\partial S_{\text{reg}}}{\partial \tilde{a}^\alpha}\,. \label{eq:Leg1}
\ee

We then use the ``conjugate" of \eqref{eq:Leg_var}, %
\be
\frac{\partial S_\text{reg}}{\partial \tilde a^\alpha}=2 \pi^2L^3 \lim_{r \to \infty}e^{-r} e^{3A}\left(L \frac{\p \mathcal{L}_{\text{bulk}}}{\p (\p_r \tilde \tau^\alpha)}+\frac{\p \cL_{\text{SUSY}}}{\p \tilde \tau^\alpha}\right) = \frac{\pi L^4}{32 G_N} \left( -g_0^2\,b^\alpha+g_\alpha \sum_\gamma g_\gamma b^\gamma\right) \,.
 \label{eq:deltaS_eta}
\ee
We insert this information in \eqref{eq:Leg1} and, using $g^2=L^{-2}$ as well, we write the free energy of this model as
\beg\label{FS3model1}
F_{S^3}=\frac{\pi L^2}{2 G_N}\left[ 1-\frac{1}{g^2}\left( g_0^2 \,\vec{c}\cdot \vec{c} - \,(\vec g\cdot\vec{c})^2\right) \right] \,.%
\eeg
We observe that the quadratic expression above involves the boundary \Kahler metric, which is real,
\be\label{eq:Kstar}
\cK_{\ak \tilde \bk}^*=\frac{1}{g^2} \left( g_0^2\delta_{\ak \tilde \bk} - g_{\ak} g_{\bk} \right)\,=\, e^a_\ak\delta_{ab} e^b_\bk,
\ee
whose frame field form from \eqref{FrameDef} is given in the final equality. 
Using this, we can recast \eqref{FS3model1} as
\be
F_{S^3} =\frac{\pi L^2}{2 G_N}\left[ 1-\vec{c} \cdot \cK^* \cdot \vec{c}\right]\,.
\ee
Next, we combine \eqref{Gotma} and \eqref{eq:eta_asymp} to express $c^\ak$ in terms of the real mass parameters:
\be
c^\alpha = \frac{i}{2} e^\ak_a \mf{m}^a\,.
\ee
Using $e^\ak_a e_\ak^b = \dk^b_a$, the free energy becomes
\be
F_{S^3} =\frac{\pi L^2}{2 G_N}\left[ 1+ \frac{1}{4} \vec{\mf m}\cdot  \vec{\mf m}\right]\,. \label{eq:FS3_eta_m}
\ee
No trace of the frame field and thus no trace of the $g_I$ couplings remains in the final formula, in agreement with the discussion in Section \ref{sec:FI_independence}.

\subsubsection*{Comparison with the conjecture}

We now reproduce the result \eqref{eq:FS3_eta_m} from the conjecture \eqref{eq:conjecture}. The only ingredients  needed are 
\begin{eqnarray}
X^I_* &=& \frac{1}{\sqrt{g^2}} (g_0,-g_i)\,,\\
(\nabla_\ak X^I)_* &=& \frac{1}{\sqrt{g^2}} (- g_\ak, g_0\dk^i_\ak)\,.
\end{eqnarray}
Using $X^I_* \eta_{IJ} X^J_* = -1$, $X^I_* \eta_{IJ}(\nabla_\ak X^J)_*=0 $ and  $(\nabla_\ak X^I)_* \eta_{IJ}(\nabla_\bk X^I)_*= - \cK^*_{\ak \bk}$, we evaluate the prepotential as instructed in \eqref{eq:conjecture}, obtaining 
\be
F_{S^3} = \frac{2\pi i L^2}{G_N} \cF(Y^I) = -\frac{\pi L^2}{2 G_N} \eta_{IJ}Y^IY^J =\frac{\pi L^2}{2 G_N}\left[ 1+ \frac{1}{4} \vec{\mf m}\cdot  \vec{\mf m}\right]\,.
\ee
This agrees with \eqref{eq:FS3_eta_m} and thus verifies the conjecture.

\subsection{$\cF=-2i \sqrt{X^0 X^1 X^2 X^3}$}
\label{sec:STU}

We now discuss the holographic free energy for the STU model which includes three vector multiplets and is described by the prepotential
\be
\cF=-2i \sqrt{X^0 X^1 X^2 X^3} \,.
\ee
The $S^3$ free energy for the specific choice of equal couplings $g_I$ was first calculated in \cite{Freedman:2013oja} without use of the prepotential.\footnote{When the couplings $g_I$ are equal, this model can be obtained as a consistent truncation of 4d $SO(8)$ gauged supergravity.  Ref.~\cite{Freedman:2013oja} used it to compute the holographic dual of the mass deformation of the $S^3$ free energy for ABJM theory \cite{Aharony:2008ug}.}  We review that calculation briefly, with emphasis on the role of the prepotential and our conjectured relation with the free energy, and afterwards we generalize this computation to the case of non-equal couplings.

\subsubsection{All couplings equal}
Following \cite{Bobev:2018uxk}, we parametrize the $Z^I$ as
\be\label{Zstu}
Z^I=\frac{1}{2\sqrt{2}} \begin{pmatrix}
	(1+\tau^1)	(1+\tau^2)	(1+\tau^3)\\
	(1+\tau^1)	(1-\tau^2)	(1-\tau^3)\\
	(1-\tau^1)	(1+\tau^2)	(1-\tau^3)\\
	(1-\tau^1)	(1-\tau^2)	(1+\tau^3)
\end{pmatrix}\,.
\ee
The formula \eqref{Kpot} gives the Kahler potential.
\be\label{Kstu}
e^\cK = \prod_{\alpha=1}^3 \frac{1}{1-|\tau^\alpha|^2}\,.
\ee
For the specific choice of equal couplings,
the superpotential \eqref{eq:W_Z}  is 
\be
W=\sqrt{2}g\left( 1+\tau^1 \tau^2 \tau^3\right) \,,
\ee
which agrees with \cite{Freedman:2013oja}.
The scalar potential \eqref{eq:V} has a supersymmetric critical point at $\tau^\alpha=\bar \tau^\alpha=0$, with AdS scale $L=1/\sqrt{2}g$. %

These are all the ingredients needed to find the solutions of the BPS equations discussed in Section~5 of \cite{Freedman:2013oja}.  In the conformally flat gauge, the solutions are 
\es{eq:bps_2013}{
\tau^\alpha &= c^\alpha \frac{1-r^2}{1+c^1c^2c^3 r^2}\,,  \qquad \tilde{\tau}^\alpha = -\frac{c^1 c^2 c^3}{c^\alpha} \frac{1-r^2}{1+c^1c^2c^3 r^2} \,, \\
e^{2A} &= \frac{4 r^2 (1+c^1 c^2 c^3)(1+c^1 c^2 c^3 r^4)}{ (1-r^2)^2 (1+c^1 c^2 c^3 r^2)^2}\,. 
}
The final result of \cite{Freedman:2013oja} is the sphere free energy, which is the Legendre transform renormalized on-shell action:
\be
F_{S^3}=\frac{\pi L^2}{2 G_N}\frac{(1-(c^1)^2)(1-(c^2)^2)(1-(c^3)^2)}{(1+c^1 c^2 c^3)^2}\,. \label{eq:FP_FS3}
\ee
In \cite{Freedman:2013oja}, this was shown to agree with the free energy in the dual SCFT, obtained by the method of supersymmetric localization in \cite{Jafferis:2011zi}.

Now we can identify the boundary real mass parameter in terms of the bulk quantities $c^\alpha$.
We make a coordinate transformation to go to the FG gauge. Its near boundary form is
\be
r \to 1-2e^{-r} +2e^{-2r} + \cO(e^{-3r})\,. \label{eq:CFtoFG}
\ee

In this gauge, the scalar fields close to the boundary behave as
\ie
\tau^\alpha &\equiv a^\alpha e^{-r}  +\ldots =\frac{4 c^\ak}{1+c^1 c^2 c^3}  e^{-r}+\ldots\\
\tilde \tau^\alpha &\equiv \tilde a^\alpha e^{-r} +\ldots =
-\frac{4 c^1 c^2 c^3}{c^\ak(1+c^1 c^2 c^3)}  e^{-r}+\ldots
\fe
A simplification occurs in our model, since the \Kahler metric is the identity at the boundary, so we can choose our frame fields to be $e_{a}^\ak = \delta_a^\ak$. This means that
\be
\mf{m}^\alpha = \frac{c^\alpha-\tilde c^\alpha}{2i} = \frac{2}{i} \frac{c^\alpha + \frac{c^1c^2c^3}{c^\alpha}}{1+c^1c^2c^3} \,. \label{eq:STU_c_m}
\ee

Now that we have all the ingredients to relate $c^\alpha$ and $\mf{m}^\alpha$, we find that
\be
\resizebox{1.1\hsize}{!}{
$F_{S^3} =  \frac{\pi L^2}{2 G_N}  \sqrt{
\left[1+\frac{i}{2}\left( \mf{m}^1+\mf{m}^2+\mf{m}^3\right)\right] 	\left[1+\frac{i}{2}\left( \mf{m}^1-\mf{m}^2-\mf{m}^3\right)\right]  \left[1+\frac{i}{2}\left(- \mf{m}^1+\mf{m}^2-\mf{m}^3\right)\right] \left[1+\frac{i}{2}\left( -\mf{m}^1-\mf{m}^2+\mf{m}^3\right)\right]  }\,. $}\label{eq:FS3_stu} 
\ee

As discussed in Section~\ref{sec:background}, the free energy of ABJM theory with trial R-charges was computed in \cite{Jafferis:2011zi}, and found to be $F_{S^3} \sim \sqrt{\Delta_1 \Delta_2 \Delta_3 \Delta_4}$, where $\Delta_j$ are the R-charges of the scalar fields transforming as bifundamentals of the $U(N) \times U(N)$ gauge group. The four terms under the square root in \eqref{eq:FS3_stu} are proportional to these R-charges when we choose pure imaginary mass parameters.

\subsubsection*{Comparison with the conjecture}
In this model the algebra needed to investigate the conjecture is quite simple, yet instructive, so we will outline the process.
Given \eqref{Zstu} and \eqref{Kstu}, we have the boundary values
\be
X^I_* = \frac{1}{2\sqrt{2}} (1,1,1,1)^T\,, \label{eq:X*stu}
\ee
and
\be
(\nabla_\alpha X^I )_*=  \frac{1}{2\sqrt{2}} \begin{pmatrix}
		1 & 1 &-1 &-1\\
		1 & -1 &1 &-1\\
		1 & -1 &-1 &1
	\end{pmatrix}_{\alpha I}\,. \label{Deq:X*stu}
\ee
Then, remembering that the boundary \Kahler metric is flat, and therefore $e_a^\ak = \delta_a^\ak$, we build the objects \eqref{eq:conjecture}
\be
Y^I = \frac{1}{4i\sqrt 2}
\begin{pmatrix}
	2i-\mf{m}^1-\mf{m}^2-\mf{m}^3\\
	2i-\mf{m}^1+\mf{m}^2+\mf{m}^3\\
	2i+\mf{m}^1-\mf{m}^2+\mf{m}^3\\
	2i+\mf{m}^1+\mf{m}^2-\mf{m}^3
\end{pmatrix} \,. \label{eq:Ystu}
\ee
When plugging this into $\mathcal{F}$, as instructed by \eqref{eq:conjecture}, we obtain exactly \eqref{eq:FS3_stu}.

\subsubsection{Unequal couplings in the STU model}

Let us now consider the case of unequal couplings $g_I$. The equations become more complicated. We will work out one case explicitly where  $g_1=g_2=g_3=g$, but $g_0 =n^2 g$ with $n^2 >0$.  Solutions for general $g_I$ are presented in Appendix~\ref{app:general_couplings}.

We parametrize the $Z^I$ as
\be\label{Zstu2}
Z^I=\frac{1}{2\sqrt{2}} 
\begin{pmatrix}
	\left(1+y^1\right)\left(1+y^2\right) \left(1+y^3\right)\\
	\left(1+y^1\right)\left(1-y^2\right) \left(1-y^3\right) \\
	\left(1-y^1 \right)\left(1+y^2\right) \left(1-y^3\right) \\
	\left(1-y^1\right)\left(1-y^2\right) \left(1+y^3\right)
\end{pmatrix}\,,
\ee
with $y^\alpha\equiv \frac{1-n}{1+n} +\tau^\alpha$. The parametrization is slightly different from the earlier one since we
 prefer to work with fields that vanish on the boundary of AdS\@.
 
 Following the usual procedure, we compute the \Kahler potential 
\be\label{Kstu2}
e^{-\cK}=\prod_{\alpha=1}^3 \left(1-|y^\alpha|^2 \right) \,,
\ee
and the superpotential
\be
W=\frac{g}{2\sqrt{2}}\left[(n^2+3)\left( 1+y^1 y^2 y^3\right)+(n^2-1)(y^1+y^2+y^3+y^1y^2+y^2y^3+y^1y^3)  \right] \,.
\ee
Computing the scalar potential, we find an extremum at $\tau^\alpha=\bar \tau^\alpha=0$, and we identify the AdS radius as $L= (\sqrt{2n}g)^{-1}$.

We insert $W$ in the BPS equations \eqref{eq:BPS_gen}, and look for zeros of the minors.
In the conformally flat gauge we find the BPS solutions
 \es{eq:sol_BPS_n}{
\tau^\alpha &= c^\alpha \frac{1-r^2}{1+v^\alpha r^2} \,, \qquad 
\tilde \tau^\alpha = \tilde c^\alpha \frac{1-r^2}{1+\tilde v^\alpha r^2} \,, \\
e^{2A} &= \frac{4 r^2 (1+w)(1+w r^4)}{(1-r^2)^2(1+wr^2)^2} \,, 
}
with
 \es{vwDefs}{
v^\alpha&=w+\frac{(n^2-1)}{4n} (w+1) c^\alpha \,, \qquad
\tilde v^\alpha=w+\frac{(n^2-1)}{4n} (w+1) \tilde c^\alpha\\
\tilde c^\alpha &= -\frac{c^1 c^2 c^3}{c^\alpha} \frac{2(n+1)^2}{8n+2(n^2-1)\sum_{\beta\neq \alpha} c^\beta +(n+1)^2(n-1) \frac{c^1 c^2 c^3}{c^\alpha}}\\
w &= \frac{c^1 c^2 c^3}{\prod_\alpha \left( \frac{4n}{(1+n)^2} +\frac{n-1}{n+1}c^\alpha \right) }\,.
 }
One can check that, for $n=1$, these formulas reduce to those in \eqref{eq:bps_2013}.

We use \eqref{eq:Sreg} in the conformally flat gauge, plug in the form of $A$ from \eqref{eq:sol_BPS_n}, integrate and take the $r_c \to 1$ limit. The result is
\be
S_{\text{reg}} = \frac{4 \pi^2 L^2}{8\pi G_N} \frac{1-w}{1+w}\,.
\ee

The next step is the the Legendre transform.  As before,  we use  the FG gauge, and the  
asymptotic coordinate transformation \eqref{eq:CFtoFG} is applicable again.
In this gauge, we use equation \eqref{eq:Leg_var}\footnote{ For $n=1$ this result agrees with (6.12) of \cite{Freedman:2013oja}, after correcting the overall minus sign missing there.}
\be
\frac{\partial S_{\text{reg}} }{\partial a^{\alpha}}=\frac{L^2 \pi }{32 G_N} \frac{(1+n)^4}{64 n^3} \left[-4 n \tilde b^\alpha+(n^2-1)  (\tilde a^\alpha)^2 +(1+n)^2 \frac{a^1 a^2 a^3}{a^\alpha} \right] \,,
\ee
as well as the variation with respect to $\tilde{a}^\alpha$. The final result for the free energy is
\be
F_{S^3} =\frac{\pi L^2}{2 G_N} \frac{16 n \prod_{\alpha} \left[ 4n+2(n^2-1)c^\alpha -(1+n)^2 (c^\alpha)^2\right] }{32 n^2 +8n(n^2-1)\sum_{\alpha} c^\alpha +2(n^2-1)^2 \sum_{\alpha} \frac{c^1 c^2 c^3}{c^\alpha}+(1+n)^3(3+n^2) c^1 c^2 c^3}\,. \label{eq:FS3_n}
\ee
For $n=1$, we reproduce the result \eqref{eq:FP_FS3}. 

 We can now use the explicit expressions \eqref{eq:sol_BPS_n} and \eqref{vwDefs} in order to relate the bulk quantities $c^\alpha$ to the boundary real mass parameter $\mf{m}_a$. Since the boundary \Kahler metric is proportional to the identity, we can choose the frame fields
\be
e^{a}_\ak = \frac{(1+n)^2}{4n} \delta^a_\ak \,. \label{eq:frame_stun}
\ee 
We go to the FG gauge, $r \to 1-2e^{-r} +2e^{-2r} + \cO(e^{-3r})$, and we extract the asymptotic behaviors of $\tau^\ak$ and $\tilde \tau^\ak$ from \eqref{eq:sol_BPS_n}. As before, the boundary real mass parameters are given by
\be
\mf{m}^a = e^a_\ak \frac{a^\ak-\tilde a^\ak}{2i}\,.
\ee
At the end of the day, we obtain again \eqref{eq:FS3_stu}. This quantity is independent of the couplings $g_I$, as expected.

\subsubsection*{Comparison with the conjecture}
As done in the previous section, we now reproduce this result using our conjecture \eqref{eq:conjecture}. Using \eqref{Zstu2} and \eqref{Kstu2}, we find
\be
X^I_* = \frac{\sqrt n}{2\sqrt2} \left(\frac{1}{n^2},1,1,1 \right)^T \,,
\ee
and
\be
(\nabla_\alpha X^I )_*=  \frac{(1+n)^2}{8\sqrt{2n}} \begin{pmatrix}
	\frac{1}{n^2} & 1 &-1 &-1\\
	\frac{1}{n^2} & -1 &1 &-1\\
	\frac{1}{n^2} & -1 &-1 &1
\end{pmatrix}_{\alpha I}\,.
\ee
Then, using \eqref{eq:frame_stun}, we build
\be
Y^I = \frac{\sqrt n}{4i\sqrt 2}
\begin{pmatrix}
	\frac{1}{n^2}\left( 2i-\mf{m}^1-\mf{m}^2-\mf{m}^3\right) \\
	2i-\mf{m}^1+\mf{m}^2+\mf{m}^3\\
	2i+\mf{m}^1-\mf{m}^2+\mf{m}^3\\
	2i+\mf{m}^1+\mf{m}^2-\mf{m}^3
\end{pmatrix} \,.
\ee

When plugging this into the prepotential $\cF(Y)$, the $n$ dependence drops out and we recover \eqref{eq:FS3_stu}, showing again agreement between our conjecture and the direct computation.

\section{Adding hypermultiplets}
\label{sec:hypers}

We now consider two examples of theories with both vector multiplets and hypermultiplets, %
where we deform the boundary SCFTs both by real mass parameters, as in the previous section, as well as by a scalar operator dual to one of the hypermultiplet scalars.
The bulk BPS equations are more complicated, and analytic solutions seem beyond reach.  So we resort to numerics, but we are still able to verify the conjecture analytically.  Both examples contain one hypermultiplet, $n_H=1$, frequently called the universal hypermultiplet, based on
the quaternionic space $\cM_V=SU(2,1)/SU(2)\times U(1)$ \cite{Cecotti:1988qn}. Its structure is reviewed in Appendix \ref{app:universal_hyper}.  The most important feature is that the moment maps are no longer constant, but are functions of the hyperscalars which are related to Killing vectors of the hypermanifold.

 The first example we treat is the special case of Section~\ref{sec:eta_vectors} with $n_V=1$ and essentially generic couplings.  We will discuss its BPS equations with the hyper added in detail. The second example is the special case of the STU model with   the specific choice of couplings of \cite{Bobev:2018wbt}.  We do not repeat the numerical analysis of that paper, but we  apply the results to our conjecture. In both models we find perfect agreement between the direct computation of the free energy and our conjecture.

We briefly mention the extra ingredients needed to specify the bulk theory with $n_H$ hypermutiplets; more details can be found in Chapter 20 of~\cite{Freedman:2012zz}. There are $4n_H$ real\footnote{For the universal hypermultiplet, the four real $q^u$ can be combined into two complex scalars $z_1,\,z_2$ and the metric $h_{uv}$ becomes a Kahler metric of dimension two.}
hyperscalar fields $q^u$ and metric tensor $h_{uv}$.  We must specify the commuting Killing vectors  $k_I^u$ which correspond to the
isometries of $\cM_H$ which are gauged by the Abelian vector fields $A_\mu^I$ of the theory.  The moment maps $\vec{P}_I$ are related to the Killing vectors.  These ingredients determine the entire bosonic action of theory. We report here the form the scalar potential:
\be
V=\left( -\frac{1}{2}\left(\Im \cN  \right)^{-1|IJ} -4  X^I \bX^J \right) \vec{P}_I \cdot \vec{P}_J  +2 \bar X^I X^J k_I^u k_J^v h_{uv}  \,. \label{eq:scalar_pot_hyp}
\ee
The structure indicates how basic quantities from the vector and hypermultiplets combine when they interact.

The AdS/CFT dual of a bulk hypermultiplet is an $\cN=2, \,d=3$ chiral multiplet plus its conjugate.  The chiral multiplet contains the following operators:  the superconformal primary, which is a scalar operator of dimension $\Delta$ and R-charge $\Delta$; a spinor operator of dimension $\Delta+1/2$ and R-charge $\Delta-1$; another scalar operator of scale dimension $\Delta+1$ and R-charge $\Delta-2$.  The conjugate operators have the same dimensions and opposite R-charges.  For more information see Section 4.2 of \cite{Binder:2021euo}.  

One can consider a supersymmetry-preserving deformation of the SCFT by adding to the Lagrangian a linear combination of the operator of dimension $\Delta+1$ and its conjugate, provided that $\Delta < 2$.  On $S^3$, such a deformation preserves the entire supersymmetry algebra $\mathfrak{osp}(2|4)$.  In examples of ${\cal N} = 2$ gauge theories with chiral multiplet charged matter, such a deformation can arise, for instance, as a relevant superpotential deformation.  In such a case, one adds to the superpotential the superconformal primary of dimension $\Delta$, and this modification of the superpotential induces a deformation of the Lagrangian by the scalar superconformal descendant of dimension $\Delta + 1$.

\subsection{The conjecture in the case of hypermultiplets}

Given the new ingredients in the $n_H\neq 0$ theory, we need to refine our conjecture. The objects we need to build are still the $Y^I$ defined in the same way as \eqref{eq:conjecture}. However, perturbing the boundary theory by the scalar superconformal descendants dual to the various hypermultiplet scalars, as discussed above, breaks some of the flavor $U(1)$ symmetries, and we end up with fewer independent real mass parameters.  We conjecture that for every hypermultiplet in the bulk theory for which we turn on such a source for the corresponding superconformal descendant, we have a further constraint
\be
Q_I Y^I = 0 \,.
\ee
Here, the $Q_I$ are the charges under the gauge fields $A_\mu^I$ of the hypermultiplet scalars dual to the superconformal primary operators.

In principle, we expect to find a supersymmetric solution if $n_H \le n_V$.
In the case $n_H = n_V$ our conjecture takes a particularly simple form. Remember that, in the case without hypermultiplets, we had the objects $Y^I$, defined in \eqref{eq:conjecture}; it is easy to show that they satisfy the property $g_I Y^I  = 1/L$, where the $g_I$ are the FI constants. This can be seen by the explicit form of the scalar potential \eqref{eq:V}, and by the critical point conditions $(\nabla_\ak W)_*=g_I  (\nabla_\ak Z^I)_*  =0$ and $V_* =-3/L^2$, together with the fact that we are working in a convention where $X^I_* = \tilde X^I_*$.

The same property holds in the case with hypermultiplets, with the only caveat that since here the moment maps are not constants, but nontrivial functions of the hyperscalars, we need to say what the $g_I$ correspond to exactly. %
We define the $g_I$ through the boundary limit of the moment maps, or equivalently the values attained for the $AdS_4$ vacuum.  In particular, at the boundary, we have $\vec{P_*}_I = g_I \vec{e}$, where $\vec e$ is some unit vector.  With this definition, we again have $g_I X^I_* = 1/L$ and $g_I (\nabla_\ak X^I)_* = 0$,\footnote{This can be seen by imposing the conditions for a critical point. See for example \cite{Freedman:2012zz}; the critical point condition corresponds to vanishing of the Goldstino, which implies the vanishing of the quantities $W_\ak^{ij}$ and ${N^i}_A$, defined in (21.40) there. The condition $W_\ak^{ij}=0$ implies, in our notation, $g_I (\nabla_\ak X^I)_*=0$. Then, from equation (21.46), with the definition (21.39), we obtain the relation $g_I X^I_* =1/L$.}
meaning that
\be
g_I Y^I = 1/L\,.
\ee
Therefore, we have $n_V+1$ variables $Y^I$ satisfying $n_H+1$ constraints. In the case $n_H = n_V$, when the charges $Q_I$ are non degenerate, the $Y^I$ are completely fixed by these constraints. We will see this in the example of Section~\ref{sec:eta_hypers}.

\subsection{$\cF=\frac{i}{4} \eta_{IJ} X^I X^J$} \label{sec:eta_hypers}

We first consider the case $n_V=1$ of the theory with prepotential $\cF=\frac{i}{4} X^I \eta_{IJ} X^J$, as described in
Section \ref{sec:eta_vectors}, but now coupled to the universal hypermultiplet \cite{Cecotti:1988qn}. For the  hypermultiplet we consider the gauging with Killing vectors
\be
k_I= (g_I \zeta + q_I \xi) \label{eq:Killing_vecs}\,,
\ee
where $\zeta$ and $\xi$ are defined in \eqref{eq:Kill_vec_basis}; they are commuting Killing vectors as needed in our Abelian theory.  

In complex coordinates, the gauge invariant scalar kinetic term in Lorentzian signature is %
\be
{\cal L}_\text{scalar kin} =  - \frac{|\p_\mu \tau|^2 }{\left( 1 - |\tau |^2\right) ^2 }- \left[  \frac{\abs{D_\mu z_1}^2 + \abs{D_\mu z_2}^2}{1 - \abs{z_1}^2 - \abs{z_2}^2}
+ \frac{\abs{ z_1^* D_\mu z_1 + z_2^* D_\mu z_2}^2}{\left( 1 - \abs{z_1}^2 - \abs{z_2}^2 \right)^2} \right] \,,
\ee
where the covariant derivatives are defined as
\ie
D_\mu z_1 &= \partial_\mu z_1 - \frac{i}{2} A_\mu^0 ( g_0 + q_0)  z_1 - \frac{i}{2}  A_\mu^1 (g_1+ q_1) z_1 \,, \\
D_\mu z_2 &= \partial_\mu z_2 - \frac{i}{2}  A_\mu^0 ( -g_0 +q_0) z_2 -\frac{i}{2}   A_\mu^1 (-g_1 + q_1)  z_2 \,. \label{eq:cov_derivatives}
\fe

Although the necessary ingredients to compute the detailed form of the potential \eqref{eq:scalar_pot_hyp} are given 
in Appendix \ref{app:universal_hyper}, this form is very complicated.
One simplification, with no loss of generality\footnote{One can check that the potential obtained with the parametrization \eqref{eq:eta_Z_param} and the Killing vectors \eqref{eq:Killing_vecs} is invariant if we rotate the the $q_I$ and the $g_I$ by the same $SO(1,1)$ rotation: $g_I \to {R_I}^J g_J$ and $q_I \to {R_I}^J q_J$, with $R = \begin{pmatrix}
		\cosh \theta & \sinh \theta\\		 \sinh \theta & \cosh \theta 
	\end{pmatrix}$.}
 is to focus on the choice of Killing vectors \eqref{eq:Killing_vecs} with $g_0=1/L$, $g_1=0$.  
The potential has a stationary point at $\tau=\bar \tau = z_i=\bar z_i=0$,
where it has value $V=-\frac{3}{L^2}$.
We now set $L=1$ in the rest of this section to simplify the notation (hence $g_0=1$). Expanding close to the stationary point, we find 
\be
V=-3-2\tau \bar \tau + (q_0 +1)(q_0-2) z_1 \bar z_1 + (q_0 -1)(q_0+2) z_2 \bar z_2+\ldots \,. \label{eq:V_expansion}
\ee
From this we can determine the correspondence between bulk fields and boundary operators: the real and imaginary parts of $\tau$  are dual  to SCFT operators $J$ and $K$ of dimension $1$ and $2$, as in the case with no hypers;  then $z_1$ and $\bar z_1$ are dual to complex operators ${\cal O}_1$ and $\bar {\cal O}_1$ of dimension $\Delta_1 = q_0+ 1$; finally $z_2$ and $\bar z_2$ are dual to complex operators ${\cal O}_2$ and $\bar {\cal O}_2$ of dimension $\Delta_2 = q_0 + 2$. 

The UV boundary theory has a $U(1)_R^{\text{UV}}$ and a $U(1)_F^{\text{UV}}$ symmetry, with $A_\mu^0$ and $A_\mu^1$ their respective bulk gauge fields.\footnote{The bulk field dual to the the UV R-symmetry can be obtained by the boundary limit of $X^I$ and $\text{Im}\, \cN_{IJ}$. See (6.80) of \cite{Lauria:2020rhc} for details.} Therefore, we see from \eqref{eq:cov_derivatives}, that the UV R-charge of $\cO_1$ is $q_0+1$ and that of $\cO_2$ is $q_0-1$. This means that $\cO_1$ is a BPS superconformal primary, since its dimension and its R-charge coincide, and $\cO_2$ is its descendant.\footnote{Another possibility from \eqref{eq:V_expansion} is to assign $\Delta_1 = 2-q_0$ and $\Delta_2 = 1-q_0$; this is consistent with $\cO_2$ being the superconformal primary and $\cO_1$ its descendant.}
From \eqref{eq:cov_derivatives}, we can read off the charges of the fields under the $U(1)_F$ symmetry in the UV. All the UV data for the hyperscalars $z_i$ are collected in Table~\ref{tab:UVCharges}.

\begin{table}[ht]
	\begin{center}
		\begin{tabular}{c|c|c|c|c}
			operator & dual field & $\Delta$ & UV R-charge $r^\text{UV}$ & UV flavor charge $f$\\
			\hline
			$J$ & $\Re \tau$ & $1$ & $0$ & $0$ \\
			$K$ & $\Im \tau$ & $2$ & $0$ & $0$ \\
			${\cal O}_1$ & $z_1$ & $q_0+ 1$ & $q_0 + 1$ & $q_1$ \\
			${\cal O}_2$ & $z_2$ & $q_0 + 2$ & $q_0 - 1$ & $q_1$ 
		\end{tabular}
	\end{center}
	\caption{UV scaling dimensions and charges.}
	\label{tab:UVCharges}
\end{table}%

The simplest strategy now is to construct a solution of the BPS equations in which only one of the two scalars $z_1,\,z_2$ is turned on. %
As per the discussion above, we should deform the boundary theory by the scalar superconformal descendant ${\cal O}_2$ (and its conjugate), which corresponds to $z_2$ (and its conjugate). Therefore we simply set $z_1=0$ and call $z_2=z$ from here onwards.  (See also \cite{Bobev:2018wbt} and \cite{Amariti:2021cpk} where the same choice was made.) This choice is consistent since
\be
\p_{z_1} V|_{z_1=\bar z_1=0}=\p_{\bar z_1} V|_{z_1=\bar z_1=0}=0\,.
\ee

Deforming the boundary theory by $\cO_2$ breaks both the UV R-symmetry and the flavor symmetry, but preserves a linear combination of the two.  
This remaining $U(1)_R^{\text{flow}}$ is the subgroup of $U(1)_R^{\text{UV}}\times U(1)_F^{\text{UV}}$ under which $\cO_2$ is invariant. This means that the charges of the fields under the $U(1)_R^\text{flow}$ symmetry are related to those of the UV symmetries by  (see Table~\ref{tab:UVCharges})
 \es{ChargeFlow}{
r = r^\text{UV} + t f, \qquad t=\frac{1-q_0}{q_1}\,,
 }
where the value of $t$ was determined from Table~\ref{tab:UVCharges} and the requirement that $\cO_2$ has vanishing charge.

In the bulk theory, we expect that in the flow toward the IR,  the gauge field  $A^R_\mu$ remains massless, while the gauge field for the broken flavor symmetry $A^m_\mu$ becomes massive.
 We now show that these gauge fields are
\be
A_\mu^R = A_\mu^0 + t A_\mu^1\,,
\qquad
A_\mu^m= A_\mu^0 (q_0 - 1) + A_\mu^1 q_1 =(q_0 - 1) \left( A_\mu^0  - \frac{1}{t} A_\mu^1  \right) \,.
\ee
First, we can rewrite the second line of \eqref{eq:cov_derivatives} as
\be
D_\mu z = \partial_\mu z - \frac{i}{2}  \left[ (q_0-1) A_\mu^0 +q_1  A_\mu^1\right]  z=\p_\mu z - \frac i 2 A_\mu^m z
\ee
and see that $z$ is uncharged under $A_\m^R$, %
which is consistent with the fact that the dual operator ${\cal O}_2$ is neutral under \eqref{ChargeFlow}. 
Second, the vector mass term along the flow is contained in the scalar kinetic term. A short calculation shows this contains
only $A_\mu^m$:
\be
D_\mu z D^\mu \tilde z \supset +\frac{1}{4} z \tilde z (A_\mu^m A^{m\,\mu})\,,
\ee
since $z(r)\,, \tilde z(r) \neq 0$ in the RG flow. 
The field $A_\m^R$ is absent and thus remains massless. %

\subsubsection{The BPS equations and their simplification}
\label{BPS}

Let us start by writing the fermion transformation rules of the theory:\footnote{These are written in Lorentzian signature in which the spinors are chiral projections of Majorana spinors.}
\ie
\delta \psi_\mu^p&= \left( \p_\mu +\frac{1}{4} \omega_{\mu}^{ab}\gamma_{ab} -\frac{i}{2} \cA_\mu\right) \epsilon^p +{\mathcal{V}_{\mu q}}^p \epsilon^q+\frac{1}{2} \gamma_\mu S^{pq} \epsilon_q  \,, \\
\delta \psi_{\mu p}&= \left( \p_\mu +\frac{1}{4} \omega_{\mu}^{ab}\gamma_{ab} +\frac{i}{2} \cA_\mu\right) \epsilon_p +{{\mathcal{V}_{\mu}}^q}_p \epsilon_q+\frac{1}{2} \gamma_\mu S_{pq} \epsilon^q \,, \\
\delta \chi_p^\alpha&=\slashed{\p} \tau^\alpha \epsilon_p+\cK^{\alpha \bar \beta} \bar{W}_{\bar \beta p q} \epsilon^q \,, \\
\delta \chi^{p \bar \alpha}&=\slashed{\p} \bar\tau^{\bar \alpha} \epsilon^p+\cK^{\beta \bar \alpha} {W_{\beta}}^{pq} \epsilon_q \,, \\
\delta \zeta^A &= \frac{i}{2} {f^{pA}}_u \slashed{\p} q^u \epsilon_p+ {\bar N_{p}}^A \epsilon^p \,, \\
\delta \zeta_{\bar A} &=-\frac{i}{2} %
{f^{qB}}_{u}\varepsilon_{qp} \rho_{B \bar A}\slashed{\p}\bar{q}^u \epsilon^p +{N^p}_A \epsilon_p \,.
\fe
The transformation rules contain the quantities
 \es{VariousDefs}{
    \cA_\mu &= \frac{i}{2}\left(\p_\mu \tau^\alpha \p_\alpha \cK-\p_\mu \bar \tau^\alpha \bar{\p}_\alpha \cK \right) \,, \\
    {\mathcal{V}_{\mu q}}^p &= ({{\mathcal{V}_{\mu}}^q}_p)^*=-{\omega_{u q}}^p \p_\mu q^u \,, \\
   S^{pq} &= (S_{pq})^*={P_I}^{pq} X^I \,, \\
   {W_{\alpha}}^{pq} &= \left( \bar W_{\bar \alpha pq}\right)^* =-P_I^{pq}\nabla_\alpha X^I  \,, \\
   {\bar N_i}^A &= ({N^i}_A)^*=-i \varepsilon_{ij} {d^{A}}_B {f^{jB}}_{u} k_I^u \bar{X}^I \,,
 }
and we have also used \eqref{eq:f_properties}. %
The moment maps $P^{pq}_I$ are expressed as rank-two symmetric tensors with $SU(2)$ fundamental indices, 
equivalent to the 3-vector notation
in \eqref{eq:moment_maps}  (see Appendix 20A of \cite{Freedman:2012zz} for more information). 

Grouping together all the equations without derivatives of the spinors $\epsilon^q, \,\epsilon_q$, and passing from Lorentzian to Euclidean signature, we have%
\be
\begin{pmatrix}
	(1+L A' e^{A-B})\delta^{p}_q & -i  L e^A S^{pq}\\
	-i  L e^A  S_{pq} & (1-L A' e^{A-B})\delta_{p}^q\\
	\cK^{\alpha \tilde \beta} \tilde W_{\tilde \beta p q} & -i e^{-B} (\tau^\alpha)' \delta_{p}^{q} \\
	i e^{-B} (\tilde{\tau}^\alpha)' \delta^{p}_{q} &\cK^{\alpha \tilde \beta} W_{\alpha}^{p q}\\
	\tilde{N}_q^A & \frac{1}{2}{f^{qA}}_u e^{-B} (q^u)'\\
	\frac{1}{2}{f^{pB}}_u \varepsilon_{pq}\rho_{B\bar A} e^{-B} (\tilde q^u)' & {N^q}_A
\end{pmatrix}
\cdot
\begin{pmatrix}
	\epsilon^q\\
	\epsilon_q
\end{pmatrix}
=0 \label{eq:all_eqs}
\ee
As already stated, we are interested in BPS solutions with $z_2=z$ and $\tilde z_2 =\tilde z$ non-vanishing, but
$z_1$ and $\tilde z_1$ set to zero. Important simplifications occur. First, we get  decoupled systems of equations for $(\epsilon_1,\epsilon^2)$ and for $(\epsilon_2,\epsilon^1)$. Further, by taking, for example, the third line of \eqref{eq:all_eqs}, solving for the $\epsilon_q$, and plugging this solution back in the last two lines, we find that
\be
\frac{z'}{\tilde z'}=\frac{z}{\tilde z}\,.
\ee
This means that $z$ and $\tilde z$ are proportional. Therefore, 
we simplify our equations henceforth by defining
 \es{XDef}{
  \zzb \equiv z \tilde z \,.
 }

We look at the system of equations for $(\epsilon^1,\epsilon_2)$.  With $z_1=\tilde z_1=0$,  moment maps vanish, except in the direction $P^3_I$. This allows us to introduce the ``superpotential''
\be
W
=-P_I^{3}Z^I=\frac{z \bar z\left(1+q_0+ q_1 \tau \right)-2}{2(1-z \bar z)}\,, \label{eq:superpot}
\ee
which is quadratically related to the scalar potential
\be
V=e^{\cK} \left( -3 W \bar W + \cK^{\tau \bar \tau} \nabla_\tau W \bar \nabla_{\bar \tau} \bar W+ 4 h^{z \bz} \p_z W \p_{\bz} \bar W\right) \,, \label{eq:V_hyper_W}
\ee
where $h^{z \bz}$ is the inverse of the quaternionic metric.

We write the simplified BPS equations in the FG gauge, which is most convenient for numerics:
\beq
\begin{pmatrix}
	1+A' e^A & 	 e^A e^{ \cK/2} W\\
	- e^A e^{ \cK/2} \tilde W &	1-A' e^A \\	
	\frac{\sqrt{1-\tau \tilde \tau }}{2} \frac{2 \tau- \zzb \left(\tau  \left(q_0+1\right)+q_1\right)}{(1-\zzb)} &\tau'\\
	\tilde \tau' &  \frac{\sqrt{1-\tau \tilde \tau }}{2} \frac{2 \tilde \tau- \zzb \left(\tilde \tau  \left(q_0+1\right)+q_1\right)}{ (1-\zzb)} \\
	\sqrt{\zzb} \frac{1-q_I \tilde Z^I}{\sqrt{1-\tau \tilde \tau}}& \frac{\zzb'}{2\sqrt{\zzb}} \\
	\frac{\zzb'}{2\sqrt{\zzb}} &   \sqrt{\zzb} \frac{1-q_I Z^I}{\sqrt{1-\tau \tilde \tau}}&
\end{pmatrix} \cdot
\begin{pmatrix}
	\epsilon^1 \\
	\epsilon_2
\end{pmatrix}=0. \label{eq:BPS_X}
\eeq
Here, $W$ is defined in \eqref{eq:superpot}, and we have omitted overall constants and common factors of $(1-\zzb)$. The same system of equations holds for $(\epsilon^1,\epsilon_2) \to (\epsilon^2,\epsilon_1)$.

Again, we require all $2\times2$ minors of \eqref{eq:BPS_X} to vanish. We start with the equations obtained from three of these minors and check later that other minors also vanish.
The determinant of the third and fifth line gives an equation for $\tau'$; the fourth and sixth line give us an equation for $\tilde \tau'$; finally, the last two lines give us an equation for $\zzb'$:
 \es{eq:BPS_system}{
\tau' &= \frac{q_1 \zzb+ \tau ((1+q_0)\zzb-2)}{4\zzb(1-\zzb)}\frac{1-\tau \tilde \tau}{q_I \tilde Z^I-1 } \zzb' \,, \\
\tilde \tau' &= \frac{q_1 \zzb+ \tilde \tau ((1+q_0)\zzb-2)}{4\zzb(1-\zzb)}\frac{1-\tau \tilde \tau}{q_I Z^I-1 } \zzb' \,, \\
(\zzb')^2 &= \frac{4(1-q_I Z^I )(1-q_I \tilde Z^I )}{(1-\tau \tilde \tau)}\zzb^2\,. 
 }
An advantage of the FG gauge is that these equations do not involve the metric. With some work, we can derive an algebraic equation for $e^{2A}$.
 Let us take the determinant of the first and sixth line, and the determinant of the second and fifth line of \eqref{eq:BPS_X};  this gives two equations for $\zzb'$:
\ie
\zzb'&=4 \frac{(1-\zzb)\zzb(1-q_I Z^I)}{\zzb(1+q_I Z^I)-2} ( A'+e^{-A}) \,, \\
\zzb'&=4 \frac{(1-\zzb)\zzb(1-q_I \tilde Z^I)}{\zzb(1+q_I \tilde Z^I)-2} ( A'-e^{-A})\,. 
\fe
By summing these two equations with the appropriate coefficients, we can eliminate $A'$, and obtain an equation for $\zzb'$ in terms of $A$ only. Then we plug this into the third line of \eqref{eq:BPS_system} and obtain 
\beq
e^{2A}=\frac{4}{ q_1^2} \frac{(1-q_I Z^I)(1-q_I \tilde Z^I)}{(\tau-\tilde \tau)^2}(1-\tau \tilde \tau)\,. \label{eq:e2A}
\eeq

The equations \eqref{eq:BPS_system} are symmetric under exchange of $\tau$ and $\tilde \tau$  (as is \eqref{eq:e2A}).  
However, there are other minors in \eqref{eq:BPS_X} which are symmetric only under the combined swap $\tau \leftrightarrow \tilde \tau$ and 
$e^A \leftrightarrow - e^A$.  We choose one solution of \eqref{eq:BPS_X} for which all minors vanish.  
There is another solution obtained by the combined swap.  The Killing spinors of these two branches are invariant under opposite choices of the $SU(2)$ factor of the isometry group $SO(4) = SU(2)_l\times SU(2)_r$ of the sphere. The computation of the free energy is the same for both solutions.

\subsubsection{Asymptotic analysis}
The goal now is to solve Eqs.~\eqref{eq:BPS_system}, and a numerical approach appears necessary. The first step is to work out the behavior of solutions in the UV and  IR\@.  We will then obtain numerical solutions which interpolate between these limits.
In order to simplify our analysis, we will focus on a specific range for $q_0$. Remember that $q_0$ determines the dimension of the CFT operators dual to $z$ and $\tilde z$,
\be
\frac{\p^2V}{\p z \p \tilde z}=(q_0+2)(q_0-1)+\ldots \,, 
\ee
and recall the  $\text{AdS}_4$ mass formula
\be
\Delta_{\pm}=\frac{3}{2} \pm \sqrt{\frac{9}{4}+m^2}\,.
\ee
It is convenient to parameterize $\Delta_+=2+q_0$ and $\Delta_-=1-q_0$, which is compatible with the bounds $\Dk_+>3/2,\,\, \Dk_-<3/2$ in the range
\be\label{q0range}
-\frac 12<q_0<\frac{1}{2}\,
\ee
within which we will work.\footnote{This also avoids the complication of mixing of the leading exponents among $\tau, \tilde\tk,$ and $\zzb$ that would happen for $q_0>1/2$.
See \eqref{eq:asymptotics}.  %
}

\subsubsection*{UV analysis}
For large $r$, we assume the following series representations for the fields:\footnote{For some special values of $q_0$, for which $\Delta_- \in \mathbb{Z}/2$, then we can also have powers of $r$ appearing in the expansion. We will avoid such special values.  %
} 
 \es{eq:asymptotics}{
\tau &= \sum_{j=1} f_{j} e^{-j r}+\sum_{n=1,j=0} f_{n,j}e^{-(2 n \Delta_-+j)r} \,, \\
\tilde \tau &= \sum_{j=1} \tilde f_j e^{-j r}+\sum_{n=1,j=0} \tilde f_{n,j}e^{-(2 n \Delta_-+j)r} \,, \\
\zzb & =\sum_{n=1,j=0} x_{n,j} e^{-(2 n \Delta_-+j)r} \,.
 }
The radial mode $e^{-\Dk_- r}$ corresponds to a source for the operators $\cO_2$ and $\tilde\cO_2.$
Notice that, naively, one would expect powers of $e^{-\Delta_+ r}$ to appear as well. Indeed, the equations of motion obtained from the bulk action allow both $e^{-\Delta_+ r}$ and $e^{+\Delta_+ r}$ behavior for the hyperscalars $z$ and $\tilde z$.
However, we see from solving the BPS equations \eqref{eq:BPS_system} that only the $e^{-\Delta_+ r}$ behavior is allowed.

We can substitute these series  in \eqref{eq:BPS_system} and solve order by order for higher mode coefficients  in terms of $f_{1}$, $\tilde f_{1}$ and $x_{1,0}$. However, we have the freedom to shift $r$; we use this to reduce the number of UV parameters by one by requiring the convenient normalization   
\beq
e^{2A}=\frac{e^{2r}}{4}+\ldots \label{eq:e2A_asymp}
\eeq
of the asymptotically AdS metric.
Examining \eqref{eq:e2A}, the condition \eqref{eq:e2A_asymp} gives us the further constraint 
 \es{f1Totildef1}{
  f_1=\tilde{f}_1+4 \frac{1-q_0}{q_1} \,.
 }  %
The first subleading terms, after shifting  $r$, are given by
 \es{eq:UV_expansion}{
  \tilde f_2 &= 2\tilde f_1 \,, \\
   f_2 &= -2\left(\tilde f_1+4 \frac{1-q_0}{q_1}\right) \,, \\
  x_{1,1} &= -2 \left(-2 q_0+\tilde f_1 q_1+2\right) x_{1,0}\,. 
  }
In practice, we compute a few orders further in order to extrapolate $\tilde f_1$ and $x_{1,0}$ from the numerical solutions to good accuracy.

\subsubsection*{IR analysis}    
We now derive the asymptotic behavior in the IR, namely where the sphere shrinks to zero size.  Let us assume this happens at some $r=r_*$, where $e^{A(r_*)}=0$.  We've already used the freedom to shift $r$ to control the UV behavior (see \eqref{eq:e2A_asymp}), and therefore we cannot shift away $r_*$. Defining $\Delta r= r-r_*$, we can solve the BPS equations \eqref{eq:BPS_system} for small $\Delta r$. The solutions contain only even powers of $\Delta r$:
\ie
\tau &= \sum_{j=0} t_j \Delta r^{2j}\,, \qquad
\tilde \tau = \sum_{j=0} \tilde{t}_j \Delta r^{2j} \,, \qquad
\zzb = \sum_{j=0} y_j \Delta r^{2j}\,.
\fe
Solving the BPS equations at leading order in $\Delta r$ gives the condition on the leading terms
\ie
 t_0&=\frac{1-q_0}{q_1} \,, \qquad 
y_0=\frac{2\tilde t_0}{q_1+\tilde t_0(1+q_0)}\,.
\fe
One can then work out further subleading orders. We see that our solutions only depend on one IR parameter, $\tilde t_0$.

\subsubsection{Numerics}
We have seen that we can solve our BPS equations as a series expansion in the UV and in the IR\@. However, we are interested in solutions which are regular all the way from the UV to the IR, and in order to see if they exist, we need to solve these equations numerically. Furthermore, we have seen that in the IR asymptotic analysis we have one free parameter, while in the UV we have two. This means that, in the case of regular solutions, the two UV parameters are not independent, and we will find numerically how they depend on the IR parameter.

When doing the numerics, we first make a shift $r \to r+r_*$; then, we would like, in principle, to integrate the BPS equations from $r=0$ to $r=\infty$. However, in practice, we choose a very small $r_{1}$ and a very large $r_2$ and numerically integrate the solutions from $r_1$ and $r_2$. Then we use the asymptotic expansions to extrapolate from $r_1$ to $0$ and from $r_2$ to $\infty$.

\begin{figure}
	\centering
	\begin{subfigure}{.5\textwidth}
		\centering
		\includegraphics[width=\linewidth]{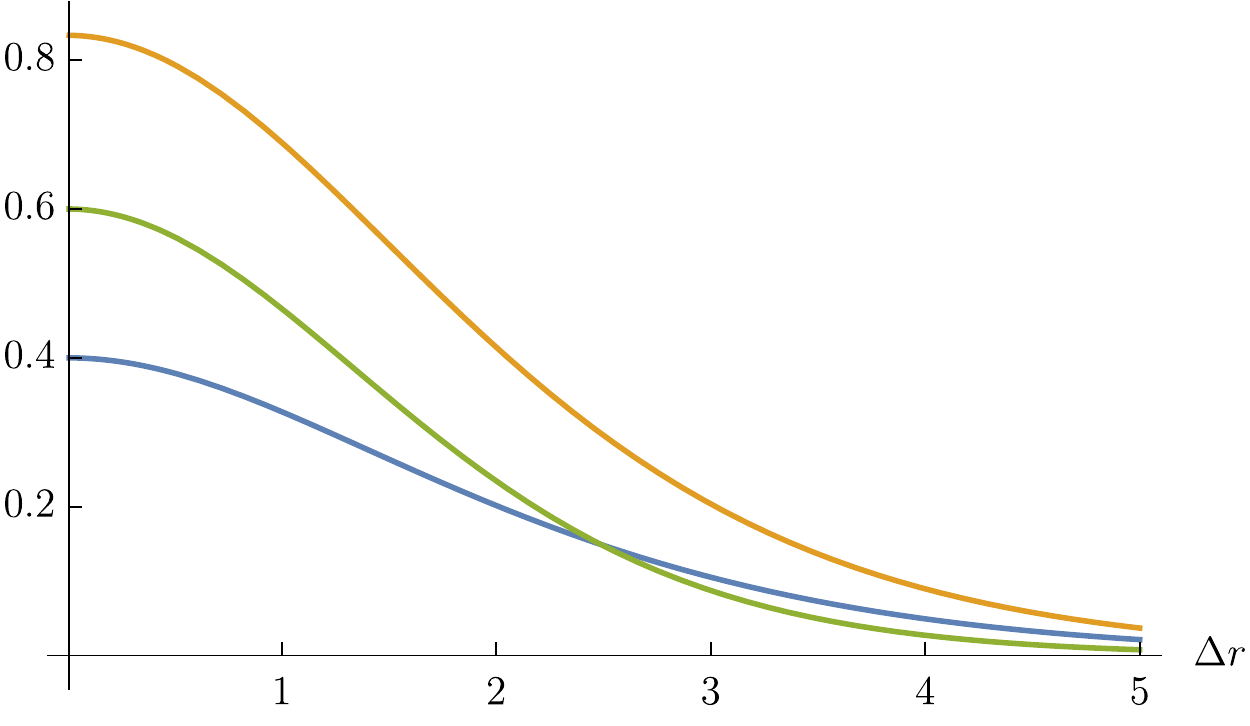}
		\caption{$q_0 = 1/3$, $q_1=4/5$, $\tilde t_0=2/5$.}
	\end{subfigure}%
	\begin{subfigure}{.5\textwidth}
		\centering
		\includegraphics[width=\linewidth]{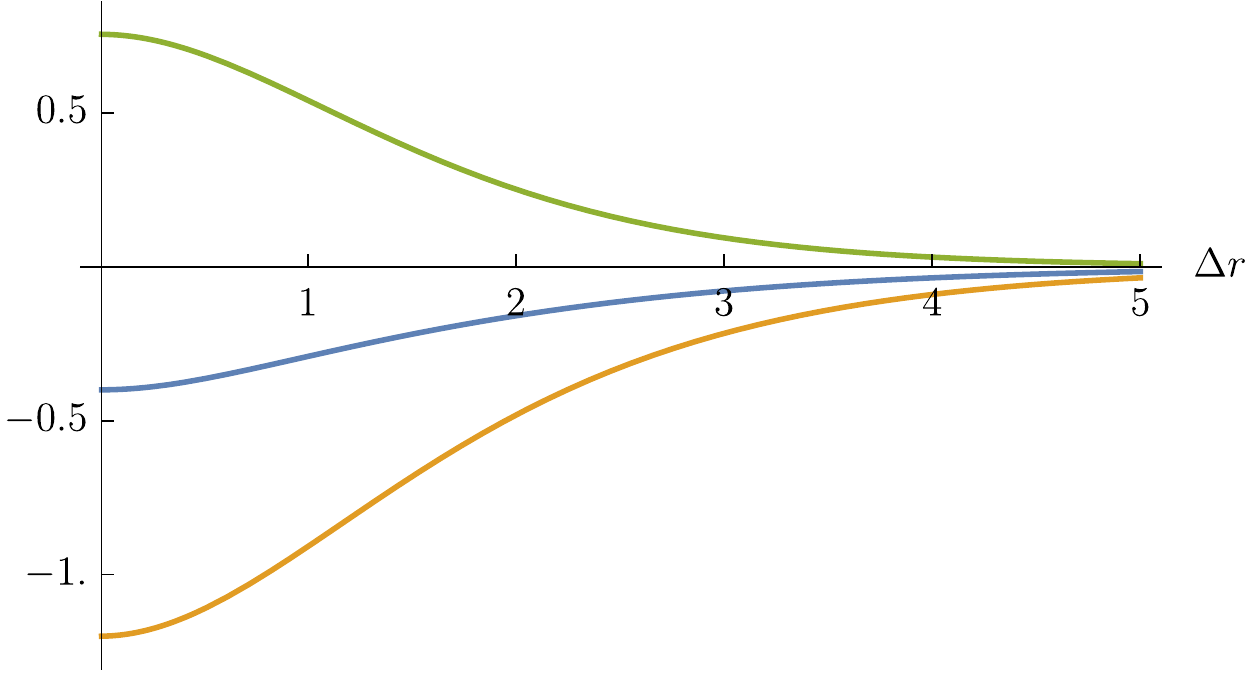}
		\caption{$q_0 = 2/5$, $q_1=-1/2$, $\tilde t_0 = -2/5$.}
	\end{subfigure}
	\caption{Example of numerical solutions for different values of $q_I$ and $t_0$.  Orange: $ \tau(r)$, blue: $ \tilde \tau(r)$, green: $\zzb(r)$.}
	\label{fig:solutions_example}
\end{figure}

We notice that regular solutions all the way to infinity exist if $\tilde t_0$ is in the range between $-\frac{q_1}{1+q_0}$ and $\frac{q_1}{1-q_0}$. We show examples of solutions in Figure~\ref{fig:solutions_example}.

\subsubsection{Computation of the free energy}

Compared to the models in Section~\ref{sec: abelian vector mults}, new counterterms may be needed to regulate the on-shell bulk action because of the presence of hypermultiplets.   These counterterms might depend on the value of $q_0$, since this determines the scaling of $\zzb$ in the UV (see \eqref{eq:asymptotics}).  It turns out that, by explicit check, the choice \eqref{eq:Ssusy} with the new superpotential \eqref{eq:superpot} makes the action finite.  One can exclude additional finite counterterms due to $\zzb$ because they do not occur at generic $q_0$.
The computation of $S_{\text{reg}}$ proceeds as in Section~\ref{sec:free_energy}.  The extra kinetic term for the hyperscalars and the different potential are absorbed via the EOM for $A$,  leading again to \eqref{eq:SA_1} (now in the FG gauge).  The  vanishing of the determinant of the first two lines of \eqref{eq:BPS_X} gives again \eqref{eq:A_W_relation}.  The net result is that $S_{\text{reg}}$ takes the same form as before, which we repeat here with $L=1$:
\be
S_{\text{reg}}=\frac{\pi}{ 2 G_N} \int_0^\infty dr\, e^{A}(A'-1)\,.
\ee
The integral converges, so it is  well suited for numerical computation.

\begin{figure}
	\centering
	\begin{subfigure}{.5\textwidth}
		\centering
		\includegraphics[width=\linewidth]{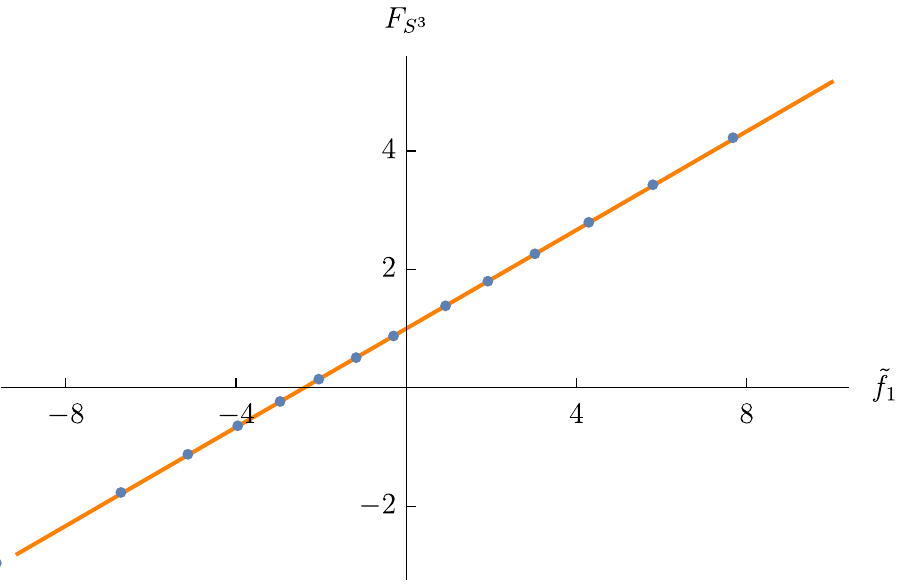}
		\caption{$q_0 = 1/3$, $q_1=4/5$, $\tilde t_0=2/5$.}
	\end{subfigure}%
	\begin{subfigure}{.5\textwidth}
		\centering
		\includegraphics[width=\linewidth]{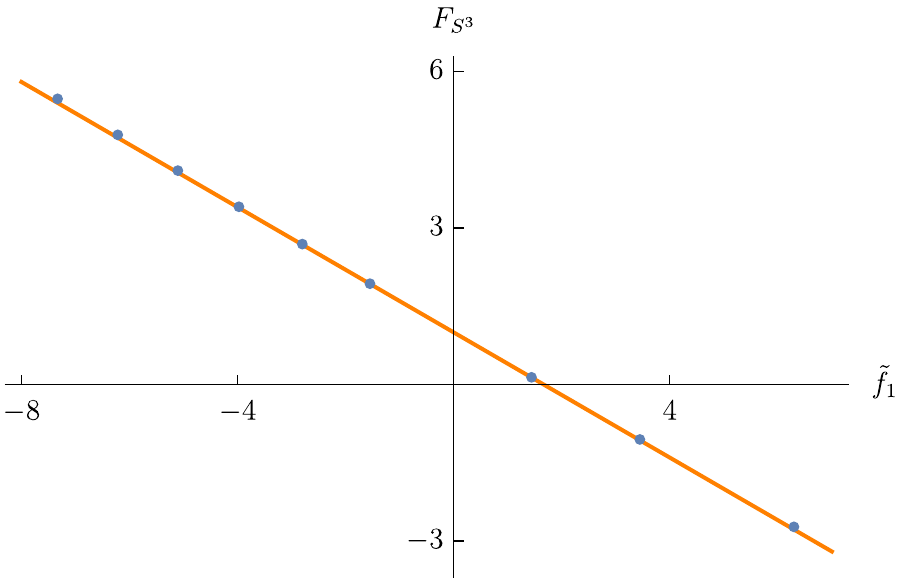}
		\caption{$q_0 = 2/5$, $q_1=-1/2$, $\tilde t_0 = -2/5$.}
	\end{subfigure}
	\caption{Value $S_\text{reg}$ in a few examples. The blue points are values of $S_\text{reg}$ computed numerically and the orange lines are the fit \eqref{eq:S_hypers}.}
	\label{fig:S}
\end{figure}

To very good numerical accuracy, and for several values of $q_0$ and $q_1$, we observe that 
\be
S_{\text{reg}}=\frac{ \pi}{ 2 G_N} \left( \frac{1-q_0}{2q_1} \tilde f_1 +1\right)\,. \label{eq:S_hypers}
\ee
This can be seen in Figure~\ref{fig:S}.

The expression \eqref{eq:S_hypers} can be checked analytically as follows.   
First, let us point out that the on-shell action $S_\text{reg}$ is naturally a function of the leading coefficients in the near-boundary asymptotic expansions \eqref{eq:asymptotics}, i.e.~$S_\text{reg} = S_\text{reg}(f_1, \tilde f_1, x_{1, 0})$, because when deriving the second order equations of motion that follow from the action, we should hold $(f_1, \tilde f_1, x_{1, 0})$ fixed. (See also the discussion in Section~\ref{LEGENDRE}.)  The expression \eqref{eq:S_hypers} equals $S_\text{reg} = S_\text{reg}(f_1, \tilde f_1, x_{1, 0})$ evaluated on supersymmetric solutions where $f_1$ is related to $\tilde f_1$ via \eqref{f1Totildef1} and $x_{1, 0}$ is still arbitrary.  However, if we do not use the supersymmetry assumption, we can calculate the partial derivatives %
\ie
\frac{\partial S_{\text{reg}}(f_1, \tilde f_1, x_{1, 0})}{\partial f_1}&=2 \pi^2 \lim_{r\to\infty} e^{-r}e^{3A}\left(\frac{\p \cL_{\text{bulk}}}{\p \tau'} + \frac{\p \cL_{\text{SUSY}}}{\p\tau}\right) =-\frac{\pi}{32 G_N} \tilde{f}_2 \,, \\
\frac{\partial S_{\text{reg}}(f_1, \tilde f_1, x_{1, 0})}{\partial \tilde f_1}&=2 \pi^2 \lim_{r\to\infty} e^{-r}e^{3A}\left(\frac{\p \cL_{\text{bulk}}}{\p \tilde \tau'} + \frac{\p
	\cL_{\text{SUSY}}}{\p\tilde \tau}\right) =-\frac{\pi}{32 G_N} {f}_2\,,
\fe

In order to consider the variation of $S_{\text{reg}}(f_1, \tilde f_1, x_{1, 0})$ with respect to $x_{1,0}$, let us remember that we defined $\zzb =z \tilde z$. If we assume, as the EOMs predict, an asymptotic behavior $z =h_1 e^{-\Delta_- r}+ h_2 e^{-\Delta_+ r}+\ldots$, and similarly for $\tilde z$, then we have $x_{1,0}= h_1 \tilde h_1$ and we find
\be
\frac{\partial S_{\text{reg}}(f_1, \tilde f_1, x_{1, 0})}{\partial x_{1, 0}} =
\frac{1}{\tilde h_1} \frac{\partial S_{\text{reg}}(f_1, \tilde f_1, x_{1, 0})}{\partial  h_1} +\frac{1}{h_1}\frac{\partial S_{\text{reg}}(f_1, \tilde f_1, x_{1, 0})}{\partial \tilde h_1}\,.
\ee
Then one can check that
\be
\frac{\partial S_{\text{reg}}(f_1, \tilde f_1, x_{1, 0})}{\partial h_1} \sim \tilde h_2\,,
\ee 
as well as its tilded version.
Remember, however, that the BPS solutions behave asymptotically as \eqref{eq:asymptotics}, meaning that $h_2 = \tilde h	_2 = 0$. Therefore we will have
\be
\left.\frac{\partial S_{\text{reg}}(f_1, \tilde f_1, x_{1, 0})}{\partial x_{1, 0}}\right|_{\text{BPS}}=0
\ee
 when evaluated on the BPS solution.
Then, when evaluated on the supersymmetric solutions parametrized by $\tilde f_1$ and $x_{1, 0}$, we can use the chain rule:
 \es{partialDers}{
\frac{\partial S_{\text{reg}}(\tilde f_1, x_{1, 0}) }{\partial \tilde f_1}= \frac{\partial S_{\text{reg}}(f_1, \tilde f_1, x_{1, 0}) }{\partial \tilde f_1}
+\frac{\partial S_{\text{reg}}(f_1, \tilde f_1, x_{1, 0})}{\partial  f_1} \frac{\partial f_1(\tilde f_1, x_{1, 0}) }{\partial \tilde  f_1} \,, \\
   \frac{\partial S_{\text{reg}}(\tilde f_1, x_{1, 0}) }{\partial x_{1, 0}}= \frac{\partial S_{\text{reg}}(f_1, \tilde f_1, x_{1, 0}) }{\partial x_{1, 0}}
+\frac{\partial S_{\text{reg}}(f_1, \tilde f_1, x_{1, 0})}{\partial  f_1} \frac{\partial f_1(\tilde f_1, x_{1, 0}) }{\partial x_{1, 0}} \,.
 }
Using \eqref{f1Totildef1} and \eqref{partialDers} and then \eqref{eq:UV_expansion}, we find
\be
\frac{\partial S_{\text{reg}}(\tilde f_1, x_{1, 0})}{\partial \tilde f_1}=-\frac{\pi}{32 G_N} \left(f_2+\tilde f_2 \right) =\frac{ \pi}{4 G_N}\frac{1-q_0}{q_1} \,, \qquad
 \frac{\partial S_{\text{reg}}(\tilde f_1, x_{1, 0})}{\partial x_{1, 0}}= 0 
\ee
This, together with the fact that $S_{\text{reg}}(0, 0) = \frac{\pi}{2 G_N}$ as evaluated in \eqref{eq:S_onshell}, gives indeed \eqref{eq:S_hypers}.

As explained previously, since the field $\tau+\tilde \tau$ is dual to a dimension one operator, in order to compute the free energy for our boundary theory, we need to do a Legendre transform of the on-shell action. The free energy is
\beg
F_{S^3}=S_{\text{reg}}-\frac 12 \left(f_1 +\tilde f_1 \right) \left(\frac{\partial S}{\partial f_1}+\frac{\partial S}{\partial \tilde f_1} \right) =S_{reg}+\frac{\pi}{64 G_N}(f_1+\tilde f_1)(f_2+\tilde f_2)=\\
=\frac{\pi}{2 G_N}\left[ 1-\left( \frac{1-q_0}{q_1}\right) ^2\right] \,.
\eeg
We see that the dependence on $\tilde f_1$ drops out, and the free energy is just a constant.

We now reinstate $L$ in the final result
\be
F_{S^3}=\frac{\pi L^2}{2 G_N}\left[ 1-\left( \frac{1-L q_0}{L q_1}\right) ^2\right] \,. \label{eq:F_num_1}
\ee

\subsubsection{Comparison to the proposal}
Since here $n_V=n_H=1$, we do not need to actually solve the bulk theory in order to compute the sphere free energy: the constraints $g_I Y^I=1/L=g_0$ and $Q_I Y^I=0$ fix uniquely $Y^I = (Y^0, Y^1)$. The quantities $Q_I$ are the UV charges of the field $z$ under the gauge potentials $A^I_\mu$, and they can be read off from the covariant derivatives appearing in the kinetic terms.  Indeed, we have
\be
D_\m z \supset -i Q_I A^I_\mu z
\ee
meaning that $Q_I =\frac 1 2 (-g_0+q_0,q_1)$.\footnote{We remind the reader that we choose $g_0=1/L$ and $g_1=0$.} Imposing the two constraints, we get
\be
Y^I = \left( 1, \frac{L q_0-1}{L q_1} \right)  \label{eq:YI_fixed}\,.
\ee
Therefore, according to our proposal, the free energy is
\beq
F _{S^3}=\frac{\pi L^2}{2 G_N}\left[ 1-\left( \frac{1-L q_0}{L q_1}\right) ^2\right]\,,
\eeq
which agrees with our direct computation \eqref{eq:F_num_1}. 

In terms of boundary charges of the chiral operator, the free energy can be written as
\be
F _{S^3}=\frac{\pi L^2}{2 G_N}\left[1-\left(\frac{r^\text{UV}}{f}\right)^2 \right]\,.
\ee
In can be checked that this is the value predicted by our conjecture for $g_1\neq 0$ as well.

\subsection{STU model with hypermultiplets}
The case of the STU model with the universal hypermultiplet was considered in \cite{Bobev:2018wbt}, for a specific choice of couplings. The authors find BPS solutions numerically and compute the free energy analytically, thanks to some simplifications similar
to those of our Section~\ref{BPS}. We will show that that their result fully agrees with our conjecture.

The choice of gauging in \cite{Bobev:2018wbt} corresponds to the Killing vectors \eqref{eq:Killing_vecs} with the $g_I=1$, for all $I = 0, \ldots, 3$, and $q_0=-3 q_i=3$, with $i = 1, 2, 3$---see Eq.~(B.23) of \cite{Bobev:2018uxk};\footnote{The scalar potential is invariant if we flip signs of both the $g_I$ and the $q_I$, so we choose the $g_I$ to be positive.} the AdS scale has been fixed to $L=\frac{1}{\sqrt{2}}$, see equation (4.1).
Notice that the conventions of our paper and \cite{Bobev:2018wbt} are different: $\tau^\ak_{\text{here}}=-\tilde z^\ak_{\text{there}}$ and $\tilde\tau^\ak_{\text{here}}=- z^\ak_{\text{there}}$. We will thus translate everything into our notation.

This theory is the bulk dual of ABJM theory perturbed by a superpotential deformation
 \es{A1Squared}{
   \Delta W \sim \Tr(T^{(1)}A_1)^2 \,,
 }
where $A_1$ %
is the one of the four bifundamental scalars, and $T^{(1)}$ is a monopole operator %
with appropriate gauge charges so as to make the superpotential deformation \eqref{A1Squared} gauge invariant.

The authors of \cite{Bobev:2018wbt} found through their numerical analysis that the profiles of the scalars $\tau_i$
change in a very simple way when the hyperscalar is included.  They introduce the parameter $x_0 =\zzb(r_*)$, where $\zzb$ is again the product of the hyperscalars $\zzb = z\tilde z$, such that the hyperscalar decouples completely when $x_0=0$.   The boundary behavior of the $\tau_i$ depends on $x_0$, i.e.
\be
\tau^\ak=a^\ak(x_0) e^{-r}+\ldots\,, \qquad \tilde\tau^\ak=\tilde a^\ak(x_0) e^{-r}+\ldots\,,
\ee
but all quantities $a^\ak$ and $\tilde a^\ak$ shift by the single function $f(x_0)$:
\be
a^\ak(x_0)=a^\ak(0)+f(x_0) \qquad \tilde a^\ak(x_0)=\tilde a^\ak(0)+f(x_0)\,. \label{eq:ax_shift}
\ee
The values $a^\ak(0),\, \tilde a^\ak(0)$ are those known from Section~\ref{sec:STU}. 
The fact that $a^\ak(x_0)-\tilde a^\ak(x_0)=a^\ak(0)-\tilde a^\ak(0)$ is independent of $x_0$, i.e.~the profile of the hyperscalars, is a major simplification which ultimately allows the authors to find the free energy analytically.

Several quantities are the same as in Section~\ref{sec:STU}: we parametrize the $Z^I$ as in \eqref{Zstu} so that the \Kahler potential is \eqref{Kstu}. The scalar potential can be computed, and there is an extremum at $\tau^\alpha=\tilde \tau^\alpha =\zzb = 0$. This means that the objects needed to construct $Y^I$, $X^I_*$ and $(\nabla_\ak X^I)_*$, are unchanged, see \eqref{eq:X*stu} and \eqref{Deq:X*stu}.
Therefore we again find
\be
Y^I = \frac{1}{4i\sqrt 2}
\begin{pmatrix}
	2i-\mf{m}^1-\mf{m}^2-\mf{m}^3\\
	2i-\mf{m}^1+\mf{m}^2+\mf{m}^3\\
	2i+\mf{m}^1-\mf{m}^2+\mf{m}^3\\
	2i+\mf{m}^1+\mf{m}^2-\mf{m}^3
\end{pmatrix} \,. \label{eq:Ystu}
\ee

We must now impose the constraint $Q_I Y^I=0$. To do this, we first discuss the charges $Q_I$ of the field $z$. As before, we determine the charges from the covariant derivative, namely
\be
D_\mu z \supset -i Q_I A_\mu^I z=\frac{i}{2} (g_I -q_I)A_\mu^I z=-i \left(A^0_\m -A^1_\m-A^2_\m-A^3_\m\right) z \,,
\ee
so that we find
\be
Q_I = (1,-1,-1,-1)\,.
\ee
Then, the constraint $Q_I Y^I =0$ gives

\be
2i + \mf m^1 + \mf m^2 + \mf m^3=0 \,. \label{eq:STU_const}
\ee

Remember that the real mass parameters are defined as
\be
\mf m^\ak =\frac{ a^\ak(x_0)-\tilde a^\ak(x_0)}{2i}=\frac{ a^\ak(0)-\tilde a^\ak(0)}{2i} \,,
\ee
where in the second equality we used the property \eqref{eq:ax_shift}. The relation between real mass parameters and IR values of the fields $c^\ak = \tau^\ak (0)$ then is the same as in the case with no hypermultiplets, \eqref{eq:STU_c_m}.

Again, the free energy can be written as \eqref{eq:FP_FS3}
\be
F_{S^3}=\frac{\pi L^2}{2G_N} \frac{(1-(c^1)^2)(1-(c^2)^2)(1-(c^3)^2)}{(1-c^1c^2c^3)^2} \,, 
\ee
where the extra constraint \eqref{eq:STU_const} translates to
\be
\frac{c^1+c^2+c^3+ c^1 c^2 +c^2 c^3+ c^3 c^1}{1+c^1 c^2 c^3}=1\,. \label{eq:1812_constraint}
\ee
This is the same constraint as (4.26) in \cite{Bobev:2018wbt}, remembering that $c_i^{\text{here}}=-\tilde c_i^{\text{there}}$.
We can also consider more general perturbations of ABJM theory, of the form 
 \es{Wp}{
  \Delta W \sim \Tr (T^{(1)} A_1)^p \,.
 }
(Only the cases $p=2, 3$ correspond to relevant perturbations.) We can find the free energy of the perturbed theories quite easily using the conjecture. The R and flavor charges of the bifundamental chiral superprimaries can be found in Table~\ref{tab:ABJM_charges}.
\begin{table}[h]
	\begin{center}
		\begin{tabular}{c|c|c|c|c}
			operator & R & $F_1$ & $F_2$ & $ F_3$\\
			\hline
			$A_1$ & $1/2$ &$1/2$ &$1/2$ &$1/2$\\
			$A_2$ & $1/2$ &$1/2$ &$-1/2$ &$-1/2$\\
			$B_1$ & $1/2$ &$-1/2$ &$1/2$ &$-1/2$\\
			$B_2$ & $1/2$ &$-1/2$ &$-1/2$ &$1/2$
		\end{tabular}
	\end{center}
	\caption{UV R and flavor charges of the chiral multiplet superprimaries.} \label{tab:ABJM_charges}
\end{table}

\begin{table}[h]
	\begin{center}
		\begin{tabular}{|c|c|c|c|c}
						\hline
			  R & $F_1$ & $F_2$ & $ F_3$\\
			\hline
 $A_0+A_1+A_2+A_3$ &$ A_0+A_1-A_2-A_3$ &$ A_0-A_1+A_2-A_3$ &$ A_0-A_1-A_2+A_3$\\
 			\hline
		\end{tabular}
	\end{center}
	\caption{Bulk gauge field duals to the UV symmetries.} \label{tab:ABJM_bulk_symm}
\end{table}

On the boundary, there are two operators, with charges
\be
\begin{tabular}{c|c|c}
	& $\Delta$ & R\\
	\hline
	$\cO_1$ & $\frac{p}{2}$ & $\frac{p}{2}$ \\
	$\cO_2$ & $\frac{p}{2}+1$ & $\frac{p}{2}-2$ \\
\end{tabular}
\ee
and F charges $F_1=F_2=F_3 = p/2$. As in previous cases, we turn on the operator $\cO_2$.  Using Table~\ref{tab:ABJM_bulk_symm}, the covariant derivative of its dual hyperscalar field $z$ can be worked out. It contains the terms
\beg
D_\m z \supset -\frac{i}{2}\left(\frac p2-2 \right) \left( A_0+A_1+A_2+A_3\right)z -\frac {ip}{4} \left(3A_0-A_1-A_2-A_3 \right) z\\
=-i(p-1)A_0z+i(A_1+A_2+A_3)z
\eeg
From this, we read off the charges $Q_I = (p-1,-1,-1,-1)$. Then the free energy is given by the usual \eqref{eq:FS3_stu}, plus the constraint $Q_I Y^I=0$, that is 
\be
2i-\mf m^1-\mf m^2-\mf m^3 = \frac{8i}{p}\,,
\ee
which fixes $Y^0 = \frac{\sqrt{2}}{p}$.

\section{Discussion}
 \label{DISCUSSION}

The main goal of this paper was to make more precise and provide additional evidence for the conjecture first explored in \cite{Hosseini:2016tor,Zaffaroni:2019dhb} that relates the mass-deformed $S^3$ free energy in holographic theories to the bulk prepotential, at the two derivative level in the bulk derivative expansion.  In the case where the bulk theory can be consistently truncated to only vector multiplet matter, this relation can be succinctly summarized in \eqref{eq:conjecture}.  In short, from the real mass parameters $\mathfrak{m}^a$ of the boundary theory one constructs quantities $Y^I$ defined in \eqref{eq:conjecture}, and then the sphere free energy is proportional to the prepotential evaluated at $Y^I$.  If the effective bulk theory contains charged hypermultiplets and one deforms the boundary theory by the operator dual to one of the hypermultiplet scalars while preserving supersymmetry, the only modification of our conjecture is that the possible real mass parameters (or equivalently the $Y^I$) obey one additional constraint for each such hypermultiplet.  

We tested our conjecture in several examples with and without hypermultiplets by explicitly constructing $SO(4)$-invariant solutions to the bulk BPS equations, evaluated their regularized on-shell action, and performed a Legendre transform corresponding to the fact that some of the bulk scalars obey alternate quantization.  In the examples with only vector multiplets, we were able to find analytical solutions to the BPS equations, but in the examples with both vector multiplets and hypermultiplets we had to resort to numerical work.  The main open question that we leave for future work is to prove our conjecture \eqref{eq:conjecture} for any bulk supergravity theory coupled to vector multiplets and hypermultiplets.  We hope that the explicit examples worked out in this paper constitute the basis for such a proof.    

Besides proving our conjecture, a natural question to ask is how to generalize it.  One class of generalizations would be to consider higher-derivative interactions in the bulk.  Given the result of \cite{Binder:2021euo} stating that the mass-deformed $S^3$ free energy is independent of bulk D-terms, $1/4$-BPS terms, and non-chiral F-terms, we expect that, in absence of real mass terms for the hypermultiplets, the $S^3$ free energy would still depend only on the prepotential of the bulk theory even when higher-derivative interactions are included.  Nevertheless, the relation between $F_{S^3}$ and the prepotential may not be as simple as equating the two quantities.  See also \cite{Bobev:2021oku} for an exploration of four-derivative terms in 4d ${\cal N} =2$ theories coupled to matter.  Assuming that $F_{S^3}$ continues to be proportional to the bulk prepotential, Ref.~\cite{Bobev:2021oku} conjectured a specific form for the leading correction to the STU model prepotential that arises in the bulk dual of ABJM theory.

Another interesting generalization of our conjecture that we hope to explore in the future is to four-dimensional ${\cal N} = 2$ SCFTs placed on a round $S^4$, which are dual to asymptotically $AdS_5$ backgrounds.  In this case too, the theory on $S^4$ can be deformed by a real mass parameter that is valued in the Cartan subalgebra of the flavor symmetry algebra of the ${\cal N} = 2$ SCFT\@.  A well-known example is the 4d ${\cal N} = 4$ super-Yang-Mills theory, which at large $N$ and large 't Hooft coupling $\lambda = g_\text{YM}^2 N$ is dual to weakly-coupled type IIB string theory on $AdS_5 \times S^5$, and which can be viewed as an ${\cal N} = 2$ theory of a vector multiplet and an adjoint hypermultiplet.  Introducing a mass parameter for the adjoint hypermultiplet yields the ${\cal N} = 2^*$ theory.  On $S^4$, the free energy of the ${\cal N} = 2^*$ theory can be calculated using supersymmetric localization \cite{Pestun:2007rz} and it is a non-trivial function of the mass $m$ \cite{Russo:2013qaa,Russo:2013kea,Russo:2019lgq,Buchel:2013id,Russo:2013sba,Bobev:2013cja}.  If it turns out that the $S^4$ free energy of the ${\cal N} = 2^*$ theory is also related to the prepotential of the effective 5d theory, then this relation is bound to be much more complicated than in the examples studied in this paper because even at leading order in the supergravity approximation, the $S^4$ free energy is a non-trivial function of the mass $m$ \cite{Bobev:2013cja}.  We hope to explore this topic in the future.

\section*{Acknowledgments}
We thanks Damon Binder for useful discussions and for collaboration on related work.  This research is supported in part by the Simons Foundation Grant No.~488653 (SSP, BZ) and by the US NSF under Grant Nos.~PHY-1620045 (DZF), PHY-2111977 (SSP), and~PHY-1914860 (BZ)\@.  

\appendix

\section{Mass spectra of abelian vector multiplet theories}
\label{app:mass spectra}

The $\cN=2$ theories discussed in Section~\ref{sec: abelian vector mults} involve only abelian vector multiplets whose interactions are governed by the prepotential $\cF(X^I)$ and the Fayet-Iliopoulos constants $g_I$.  The general formulas of Section~21.3 of \cite{Freedman:2012zz} simplify greatly for these theories,
since there are no hypermultiplets, and Killing vectors $k_I^\ak$ vanish because all scalars are uncharged.

Some theories have holographic duals. In this case each abelian vector multiplet
$(A_\mu, \chi_1,\chi_2, \tau)$ is dual to a conserved current multiplet containing the operators %
$(J, \xi_1,\xi_2, j^a,K)$ with conformal dimensions $\Dk  = (1,3/2,3/2,2,2)$ and $D_aj^a=0$.  The bulk field masses that correspond to these values of $\Dk$ are $m^2 =0$ for the vector and spinors and $m^2=-2/L^2$ for the scalar.  In this Appendix we derive these mass spectra using only the properties of the bulk theory, independent of holography.

In the first step we use only $\cN=1$ information to show that scalars have $m^2 =-2/L^2$ if two conditions are satisfied:
\begin{enumerate}
	\item The boundary values of the scalars $\tau^\ak_*$ correspond to a SUSY critical point (c.p.)~where
	\be\label{eq:SUSYcp}
	\nabla_\ak W(\tk)|_{\tk =\tk_*} \equiv (\partial_\ak +\cK_\ak)W(\tau)|_{\tk =\tk_*}=0 \,.
	\ee
	\item The fermion mass matrix, given in detail in (18.17) of \cite{Freedman:2012zz}, also vanishes at the critical point
	\be \label{eq:fmass}
	m_{ab} \equiv e^{\cK/2}\nabla_\ak\nabla_\bk W|_{\tk =\tk_*}=0\,.
	\ee
\end{enumerate}

We start with the standard $\cN=1$ potential \eqref{eq:V}
\be
V=e^\cK (-3 W \bW+\cK^{\gk \bd} \nabla_\gk W \nabla_\bd \bW)\,.%
\ee
We apply two derivatives and push them to the right,  noting that $\pa_\ak (e^\cK \cdots) = e^\cK \nabla_\ak (\cdots) $:
\be
\pa_\bb\pa_\ak V =e^\cK \nabla_\bb\nabla_\ak (-3 W \bW+\cK^{c \bd} \nabla_cW \nabla_\bd \bW)\,.
\ee
Let's consider the two terms separately. The first term is easy to take care of:
\be 
\pa_\bb\pa_\ak V_1 =-3 e^\cK(\nabla_\ak W\nabla_\bb\bW + \cK_{\ak\bb} W\bW)|_{\tk =\tk_*} =-3e^\cK \cK_{\ak\bb} W\bW\,.
\ee
Notice that the K\"ahler metric appears.  When working with the second term, we have to push a little harder. The derivative $\nabla_\ak$ acts covariantly
with the metricity property 
$\nabla_\ak (\cK^{\gk\bd} \cdots)  =\cK^{\gk\bd}(D_a+K_a) (\cdots)$, where $D_a$ is the usual Riemannian covariant derivative
specialized to Kahler manifolds.  We expand out  $\nabla_\bb\nabla_\ak$ for clarity, and write
\begin{eqnarray}
	\partial_\bb\partial_\ak V_2 &=& e^\cK (\partial_\bb+\cK_\bb)\cK^{\gk\bd}[((D_\ak+\cK_\ak)\nabla_\gk W)\nabla_\bd\bW +\nabla_\gk W \cK_{\ak\bd}\bW]\\  \nonumber
	&=& e^\cK\big\{ \cK^{\gk\bd}[K_{\ak\bb}\nabla_\gk W\nabla_\bd \bW + (\nabla_\ak\nabla_\gk W)(\nabla_\bb\nabla_\bd \bW)] +R_{\ak\bb}{}^{\gk\bd}\nabla_\gk W \nabla_\bd\bW\\ \nonumber
	&+&\nabla_\ak W \nabla_\bb\bW + K_{\ak\bb} W\bW\big\}|_{\tk =\tk_*} = e^\cK \cK_{\ak\bb} W\bW\,.
\end{eqnarray}
The 3 terms in the second  line come from the first term in $(\cdots)$ in the line above, while the second term in $(\cdots)$ gives the 2 terms in the third line.  The curvature tensor of the Kahler manifold comes from the commutator $[D_\ak,D_\bb]$.
All terms except the last one vanish when  the conditions at the SUSY c.p.~are imposed. Finally,  we combine the results for $V_1$ and $V_2$ to obtain (at the c.p.)
\be
\pa_\bb\pa_a V = -2e^\cK \cK_{a\bb} W\bW |_{\tau=\tau^*} = -2\cK_{a\bb}/L^2\, .
\ee
By a similar calculation, we find
\be
D_\ak\partial_\bk V|_{\tk =\tk_*} = e^\cK \big[ K^{\gk\bd} (\nabla_\ak\nabla_\bk\nabla_\gk W) \nabla_\bd \bW -(\nabla_\ak\nabla_\bk W)\bW\big]|_{\tk =\tk_*}=0\,,
\ee
showing that there are no exotic contributions to scalar masses.

The kinetic and mass Lagrangian for the scalar fields then reads, in Lorentzian signature,
\be
\cL = \sqrt{-g}\cK_{\ak\bb} \left[ -g^ {\m\n}\partial_\mu \tk^\ak\partial_\nu\tau^\bb + \frac{2}{L^2}\tau^\ak\tau^\bb\right] \,.
\ee
Since $\cK_{\ak\bb}$, evaluated at the critical point, appears as an overall factor, the scalar mass is indeed $m^2 =-2/L^2$.

We now bring in $\cN=2$ information which shows that the fermion mass vanishes at a SUSY critical point, so that \eqref{eq:fmass} is actually a consequence of \eqref{eq:SUSYcp}.
The fermion mass term in $\cN =2$ supergravity is given in (21.38-.39) of \cite{Freedman:2012zz} with $C_{\ak\bk\gk}$ defined in (20.236):
\be\label{eq:n2f}
\cL_m =-\frac12 m^{ij}{}_{\ak\bk} \,\bar\chi^\ak_i\chi^\bk_j, \qquad  m^{ij}{}_{\ak\bk}=\frac12 P_I^{ij} C_{\ak\bk\gk}\cK^{\gk\bd}\nabla_\bd\bar X^I\,.
\ee
Another $\cN=2$ relation we need is (20.191) (which is proven above (5.75) of \cite{Lauria:2020rhc}):
\be\label{20191}
\nabla_\bk\nabla_\ak X^I = C_{\ak\bk\gk}\cK^{\gk\bd}\nabla_\bd\bar X^I\,.
\ee
We must adapt this information to the $\cN=1$ truncation   we have been using. It is a truncation because one of the two
sets of gauginos $\chi_i$ is dropped.  Note that the moment maps $P_I^{ij}$ are related to the FI constants by $P_I^{ij}=g_I \sk_1^{ij}$ where $\sk_1^{ij}$ is the standard Pauli matrix, with eigenvalues $\pm 1$. We will keep the fermion with positive eigenvalue in our truncation. Thus \eqref{eq:n2f} is replaced by
\be\label{eq:n1f}
\cL_m = -\frac12 m_{\ak\bk} \,\bar\chi^\ak\chi^\bk, \qquad m_{\ak\bk}= C_{\ak\bk\gk}\cK^{\gk\bd}\nabla_\bd(g_I\bar X^I)= e^{\cK/2}C_{\ak\bk\gk}\cK^{\gk\bd}\nabla_\bd \bW.  
\ee
In the last equality we used $g_IX^I = e^{\cK/2}W$, see \eqref{eq:W_Z}.  We contract \eqref{20191} with $g_I$, and rewrite the result as
\be\label{n1n2}
e^{\cK/2}\nabla_\bk\nabla_{\ak} W =  e^{\cK/2}C_{\ak\bk\gk}\cK^{\gk\bd}\nabla_\bd \bW.
\ee
This equation shows that the $\cN=1$ and $\cN=2$ fermion mass matrices are the same. Then \eqref{eq:n1f} shows that this mass matrix vanishes at a SUSY critical point.
Finally we point out that gauge fields are massless in a theory of abelian vector multiplets without hypermultiplets. The $\tau^\ak$ fields are gauge neutral, so no opportunity to generate a vector mass term arises. 

\section{Asymptotic analysis}
\label{ASYMPTOTIC}

In this Appendix, we consider an ${\cal N} = 2$ supergravity theory coupled to $n_V$ vector multiplets and determine which scalar fields obey which boundary conditions.  Doing so requires that we examine the fluctuations of the fields around an AdS background where the vector fields and spinors all vanish, and the scalar fields are constant $X^I = X^I_*$, $\tau^\alpha = \tau^\alpha_*$, etc.  In such a background, the SUSY variations of the bosonic fields vanish automatically.  The SUSY variations of the fermionic fields vanish provided that
 \es{Conditions}{
  \delta \psi_\mu^i &= D_\mu \epsilon^i + \frac 12 \gamma_\mu \tau_{3 ij} g_I \bar X^I_* = 0 \,, \\
  \delta \chi_i^\alpha &= -g_I \tau_{3 ij}  g^{\alpha \bar \beta}_* \left( \bar \nabla_{\bar \beta} \bar X^I \right)_*  \epsilon^j = 0 \,,
 }
where, as in the main text, the subscript ${}_*$ means that the quantities are evaluated at the constant values of the fields corresponding to the AdS solution.  In addition to \eqref{Conditions}, we also have the complex conjugate equations.

As mentioned in the main text, we take $g_I \bar X^I_*  = g_I X^I_*= 1/L$ so that the first equation in \eqref{Conditions} reduces to the Killing spinor equation
 \es{KSEqAppendix}{
  D_\mu \epsilon^i = - \frac{1}{2L} \gamma_\mu \tau^{ij} \epsilon_j \,, \qquad
   D_\mu \epsilon_i = - \frac{1}{2L} \gamma_\mu \tau_{ij} \epsilon^j  \,.
 }
From the second equation in \eqref{Conditions} and its complex conjugate, we have the SUSY conditions
 \es{SUSYCond}{
   g_I  \left( \nabla_{ \alpha} X^I \right)_* = g_I  \left( \bar \nabla_{\bar \alpha} \bar X^I \right)_* = 0 \,.
 }

Even though we are interested in solutions in Euclidean signature, the question of whether or not scalar fields obey alternate quantization can be settled using Poincar\'e coordinates.  we thus take the frame to be $e^a = e^{r/L} dx^a$ and $e^3 = dr$, for $a =0, 1, 2$.  We have $\omega^{a3} = e^a / L$ and $\omega^{ab} = 0$.  The solution of the KS equation \eqref{KSEqAppendix} corresponding to the Poincar\'e supercharges\footnote{We do not need the Killing spinors for superconformal charges, since those symmetries are broken in a mass-deformed theory.} is
 \es{KSSolution}{
  \epsilon^i = e^{r/2L} \eta^i_+ \,, \qquad
   \epsilon_i = e^{r/2L} \eta_{i+} \,,
 }
where $\eta^i$ and $\eta_i$ are constant spinors obeying the radiality conditions
 \es{Radiality}{
  \eta^i_+ = - \gamma_3 \tau^{ij} \eta_{j+} \,, \qquad
    \eta_{i+} = - \gamma_3 \tau_{ij} \eta^j_+ \,.
 }

Let us now study how SUSY acts on the coefficients of the boundary expansion of the linearized fluctuations of the fields around the AdS solution.  (For simplicity, $L=1$.)   We have 
 \es{Expansion}{
  \tau^\alpha &= \tau^\alpha_* + \tau^\alpha_1 e^{-r} + \tau^\alpha_2 e^{-2r} + \cdots \,, \qquad
   X^I = X^I_* + X^I_1 e^{-r} +  X^I_2 e^{-2r} + \cdots \,, \\
  \chi_i^\alpha &= \chi_{i 3/2}^\alpha e^{-3r/2} +  \chi_{i 5/2}^\alpha e^{-3r/2} + \cdots \,, \\
  A_a & = A_{a 1} e^{-r} + A_{a 2} e^{-2r} + \cdots \,, \\
  \psi^i_a &= \psi^i_{a 1/2} e^{-r/2}  + \psi^i_{a 3/2} e^{-3r/2}  + \cdots \,,
 } 
where the gauge field is written with frame indices.  For the fermions, we can split the coefficients based on their radiality as in
 \es{chiRad}{
  \chi_{i 3/2 \pm }^\alpha =  \mp \gamma_3 \tau_{ij} \chi^{j\alpha}_{3/2\pm } \,, \qquad \text{etc.}
 }
SUSY acts on this asymptotic data via the transformation rules 
 \es{SUSYVarsUse}{
  \delta \tau^\alpha &= \frac 12 \bar \epsilon^i \chi_i^\alpha \,, \\
  \delta A_\mu{}^I &= \frac 12 \varepsilon^{ij} \bar \epsilon_i \gamma_\mu \chi_j^\alpha (\nabla_\alpha X^I)_* + \frac 12 \varepsilon_{ij} \bar \epsilon^i \gamma_\mu \chi^{j\alpha} (\nabla_\alpha \bar X^I)_* 
   + \varepsilon^{ij} \bar \epsilon_i \psi_{\mu j} (X^I)_* +  \varepsilon_{ij} \bar \epsilon^i \psi^j_{\mu} (\bar X^I)_* \,,
 }
as well as their complex conjugates.  The $\epsilon^i,\,\epsilon_i$ are the Killing spinors of \eqref{KSSolution}.

Before proceeding, we note that we are  interested in the transformation of the $n_V$ vector multiplet photons, which couple to the $\chi_\alpha$, rather than the gravi-photon,  
which couples only to the gravitino. Near the boundary these properties of the couplings can be seen by contracting the $  \delta A_\mu{}^I$ transformation rule with $g_I$ and using $g_I \bar X^I_*  = g_I X^I_*= 1/L$ together with \eqref{SUSYCond}:
\es{grvphot}{
\delta g_I A_\mu{}^I  =[\varepsilon^{ij} \bar \epsilon_i \psi_{\mu j} +  \varepsilon_{ij} \bar \epsilon^i \psi^j_{\mu}]/L\,.
}
Likewise, one can contract $  \delta A_\mu{}^I$ with any of the $n_V$ independent sets of coefficients $h_I$ which satisfy $h_I(X^I)_* =0$. 
These vector multiplet photons $h_I A_\mu{}^I$ couple only to their fermionic partners.  Thus we drop the gravitino in the following with the understanding that we are considering a contraction with a set $h_I$.

 Returning to \eqref{SUSYVarsUse}, we use the first equation there, expanded at large $r$ as in \eqref{Expansion}, to obtain
 \es{deltazCoeff}{
  \delta \tau^\alpha_1 
   &= \frac 12 \bar \eta^i_+ \chi^\alpha_{i 3/2} \,, \qquad
     \delta \tau^\alpha_2 
   = \frac 12\bar \eta^i_+ \chi^\alpha_{i 5/2} \,,  \\
   \delta \bar \tau^\alpha_1 
   &= \frac 12\bar \eta_{i+} \chi^{i\alpha}_{3/2} \,, \qquad
     \delta \tau^\alpha_2 
   = \frac 12\bar \eta_{i+} \chi^{i\alpha}_{ 5/2} \,,  \qquad \text{etc.}
 }
 We have $\bar \eta^i_+ \chi^\alpha_{i n + } = - \bar \eta_{i+} \chi^{i \alpha}_{n+} $ and $\bar \eta^i_+ \chi^\alpha_{i n -} = \bar \eta_{i+} \chi^{i \alpha}_{n-} $.  Thus, if we write $\tau^\alpha = A^\alpha + i B^\alpha$, then
  \es{deltaABCoeff}{
   \delta A^\alpha_1 &= \frac 12 \bar \eta^i_+ \chi^\alpha_{i 3/2 - } \,, \qquad 
     \delta A^\alpha_2 = \frac 12 \bar \eta^i_+ \chi^\alpha_{i 5/2 - } \,, \\
   \delta B^\alpha_1 &= -\frac i2 \bar \eta^i_+ \chi^\alpha_{i 3/2 + } \,, \qquad 
     \delta B^\alpha_2 = -\frac i2 \bar \eta^i_+ \chi^\alpha_{i 5/2 + } \,, \qquad \text{etc.,}    
  }
where $A_n^\alpha$ and $B_n^\alpha$ are the coefficients of $e^{-n r}$ in the boundary expansion of $A^\alpha$ and $B^\alpha$.
 
From the second equation in \eqref{SUSYVarsUse}, we get
 \es{deltaAaCoeffs}{
  \delta A_{a 1}{}^I &= \frac 12 \varepsilon^{ij} \bar \eta_{i+} \gamma_a \chi_{j 3/2}^\alpha (\nabla_\alpha X^I)_* + \frac 12 \varepsilon_{ij} \bar \eta^i_+ \gamma_a \chi^{j\alpha}_{3/2}  (\nabla_\alpha \bar X^I)_*    \,, \\
   \delta A_{a 2}{}^I &= \frac 12 \varepsilon^{ij} \bar \eta_{i+} \gamma_a \chi_{j 5/2}^\alpha (\nabla_\alpha X^I)_* + \frac 12 \varepsilon_{ij} \bar \eta^i_+ \gamma_a \chi^{j\alpha}_{5/2}  (\nabla_\alpha \bar X^I)_*  \,.
 }
Note that $\varepsilon^{ij} \bar \eta_{i+} \gamma_a \chi_{j 3/2 \pm}^\alpha = \pm \varepsilon_{ij} \bar \eta^i_{+} \gamma_a \chi_{3/2 \pm}^{j\alpha}$, so
 \es{deltaAaCoeffsAgain}{
   \delta A_{a 1}{}^I &= \frac 12 \varepsilon^{ij} \bar \eta_{i+} \gamma_a \chi_{j 3/2 + }^\alpha (\nabla_\alpha X^I + \nabla_\alpha \bar X^I)_* + \frac 12 \varepsilon^{ij} \bar \eta_{i+} \gamma_a \chi_{j 3/2 - }^\alpha (\nabla_\alpha X^I - \nabla_\alpha \bar X^I)_* \,, \\
     \delta A_{a 2}{}^I &= \frac 12 \varepsilon^{ij} \bar \eta_{i+} \gamma_a \chi_{j 5/2 + }^\alpha (\nabla_\alpha X^I + \nabla_\alpha \bar X^I)_* + \frac 12 \varepsilon^{ij} \bar \eta_{i+} \gamma_a \chi_{j 5/2 - }^\alpha (\nabla_\alpha X^I - \nabla_\alpha \bar X^I)_* \,.
 }

In the boundary expansion of the gauge field, we know that the leading coefficient $A_{a1}^I$ should be interpreted as the source for the dual conserved current on the boundary:
 \es{ASource}{
  A^I_{a, \text{source}} = A_{a1}^I \,.
} 
From the first line of \eqref{deltaAaCoeffsAgain} and the fact that sources transform into sources, we have 
 \es{deltaA1Generic}{
  A^I_{a, \text{source}}  = \varepsilon^{ij} \bar \eta_{i+}\gamma_a \chi^I_{j, \text{source}} \,,
 }
where we identify
 \es{BCFerm}{
  \chi^I_{j, \text{source}} \equiv  \chi_{j 3/2 + }^\alpha (\nabla_\alpha X^I + \nabla_\alpha \bar X^I)_* + \chi_{j 3/2 - }^\alpha (\nabla_\alpha X^I - \nabla_\alpha \bar X^I)_*  \,.
 }
The quantities $\chi^I_{j, \text{source}}$ then represent the sources for the dimension $3/2$ fermionic operators dual to the massless bulk fermions.  Note that $g_I \chi^I_{j, \text{source}} = 0$, so there are only $n_V $ linearly independent $\chi^I_{j, \text{source}}$.

We have mentioned that in the boundary theory each conserved current multiplet contains a dimension $2$ scalar (dual to a regularly quantized bulk field) and a dimension $1$ scalar operator (dual to an alternately quantized field).  The source ${\cal B}^I_\text{source}$ for the dimension $2$ scalar operator must be a linear combination of the leading coefficients $A_1^I$ and $B_1^I$ and must obey the property that SUSY transformations take it to the fermion sources in \eqref{BCFerm}
 \es{deltacalB}{
  \delta {\cal B}^I_\text{source} =\bar \eta^i_+ \chi^I_{i, \text{source}} \,. 
 }
 Examining \eqref{deltaABCoeff}, we identify
 \es{BCScalars}{
   \delta {\cal B}^I_\text{source} = i B_1^\alpha (\nabla_\alpha X^I + \nabla_\alpha \bar X^I)_* +A_1^\alpha (\nabla_\alpha X^I - \nabla_\alpha \bar X^I)_* =  \Im ( \tau_1^\alpha (\nabla_\alpha X^I)_* ) \,.
 } 
Thus, it is the field $ {\cal B}^I = \Im ( (\tau^\alpha - \tau^\alpha_*) (\nabla_\alpha X^I)_*)  $ that should have regular boundary conditions, while $ {\cal A}^I = \Re ( (\tau^\alpha - \tau^\alpha_*) (\nabla_\alpha X^I)_*) $ should obey alternate quantization.  The fields ${\cal A}^I$ and ${\cal B}^I$ are not independent because the condition \eqref{SUSYCond} implies $g_I {\cal A}^I = g_I {\cal B}^I = 0$.  There are of course only $n_V$  linearly-independent fields of each kind.

\section{The STU model with general couplings}
\label{app:general_couplings}
The solutions to the BPS equations become a bit more complicated  in the STU model when we choose all couplings to be different; however, we're still able to find a solution analytically. We change slightly parametrization for the $Z^I$, because we want the scalar fields to vanish at the boundary
\be
Z^I=\frac{1}{2\sqrt{2}} 
\begin{pmatrix}
	\left(1+y^1\right)\left(1+y^2\right) \left(1+y^3\right)\\
	\left(1+y^1\right)\left(1-y^2\right) \left(1-y^3\right) \\
	\left(1-y^1 \right)\left(1+y^2\right) \left(1-y^3\right) \\
	\left(1-y^1\right)\left(1-y^2\right) \left(1+y^3\right)
\end{pmatrix}
\ee
with
\beg
y^\alpha=\Omega^\alpha + \tau^\alpha\,.
\eeg
For convenience, we use the notation
\beg
\Omega_a=\frac{  \frac{1}{L^2}-\frac{(\xi^+_a)^2+(\xi^-_a)^2}{2}}{\xi^+_a\xi^-_a}\,,\\
\xi_a^\pm=\sqrt{g_0 g_a}\pm\frac{\sqrt{g_1 g_2 g_3}}{\sqrt{g_a}}\,,\\
L^2=\frac{1}{2\sqrt{g_0 g_1 g_2 g_3}}\,.
\eeg
The solution to the BPS equations for the scalar fields and the metric is 
\beg
\tau^\alpha=c^\alpha \frac{1-r^2}{1+v^\alpha r^2}\\
\tilde \tau^\alpha=\tilde c^\alpha \frac{1-r^2}{1+\tilde v^\alpha r^2}\\
e^{2A}=4r^2\frac{(1+w)(1+w r^4)}{(1-r^2)^2(1+w r^2)^2}\,, \label{eq:sol_gen_case}
\eeg
where we defined the constants
\beg
v^\ak=w+c^\ak\frac{L^2}{2}\xi^+_\ak \xi^-_\ak (1+w)\\
\tilde{v}^\ak=w+\tilde{c}^\ak\frac{L^2}{2}\xi^+_\ak \xi^-_\ak (1+w)\\
\tilde{c}^\ak=\frac{c^1 c^2 c^3}{c^\ak} \frac{(\xi^+_1 \xi^+_2 \xi^+_3)^2}{(\xi^+_\ak)^4}\left[-\frac{2}{L^2}-\sum_{j\neq \ak} c^j \xi_j^+ \xi_j^- -\frac{1}{2}\frac{c^1 c^2 c^3}{c^\ak}\frac{\xi^+_1 \xi^+_2 \xi^+_3}{(\xi^+_\ak)^2}\left(g_0+g_\ak -\sum_{j\neq \ak}g_j \right)  \right]^{-1}\\
w=\frac{L^6}{8} \left(\prod_{j} \xi^+_j \xi^-_j\right) \frac{\sum_I \left( g_I +\frac{1}{2L^2} \frac{1}{g_I}\right) }{\sum_I \left( g_I -\frac{1}{2L^2} \frac{1}{g_I}\right)} \frac{c^1 c^2 c^3}{\prod_j \left( 1+\frac{L^2}{2}\xi_j^+ \xi_j^- c^j\right) } \label{eq:sol_expl}
\eeg
where sum and products over lower case letters are intended to be from $1$ to $3$, while upper case ones are from $0$ to $3$.

The computation of $S_{\text{reg}}$ follows from Section~\ref{sec:STU}, and is
\beq
S_{\text{reg}}=\frac{\pi L^2}{2 G_N} \frac{1-w}{1+w}\,.
\eeq
Notice that in the case of all couplings equal, $w$ reduces to $c^1 c^2 c^3$, as can be seen by \eqref{eq:sol_expl}. In this case our result agrees with formula (6.19) of \cite{Freedman:2013oja}. It can also be seen that the case $g_0=n^2 g$, $g_1=g_2=g_3=g$ agrees with \eqref{eq:FS3_n}.

Then we need to perform the Legendre transform. We find
\be
\frac{\partial S_\text{reg}}{\partial a^\alpha}=-\frac{L^2 \pi}{32 G_N}  \left(\frac{\Omega_\alpha \tilde (a^\alpha)^2}{(1-\Omega_\alpha)^3} +\frac{ \tilde b^\alpha}{(1-\Omega_\alpha)^2}-\frac{L^6}{8} \left( \prod_j \xi^+_j\right) ^2 \frac{a^1 a^2 a^3}{a^\alpha}\right) \,.
\ee

Putting everything together, we get that the free energy is
\be
F_{S^3}=\frac{4 \pi}{G_N} \prod_j \left( \frac{2}{L^2}+2\xi_j^+ \xi_j^-  c^j  -(\xi_j^+ c^j)^2\right) \left[\frac{8}{L^{4}}+\frac{4}{L^2} \sum_j \xi_j^+ \xi_j^- c^j +2 \sum_j \frac{\prod_i \xi_i^+ \xi_i^- c^i}{\xi_j^+ \xi_j^- c^j} +\prod_j \xi_j^+ \sum_I g_I\right] ^{-1}\,. \label{eq:FS3_alldiff}
\ee

Then, using 
\be
\frac{a^\ak-\tilde a^\ak}{2i} = -2i \left( \frac{c^\alpha}{1+v^\ak} - \frac{\tilde c^\alpha}{1+\tilde v^\ak} \right) \,,
\ee
as well as
\be
\cK^*_{\ak \tilde \beta} = \frac{1}{(1-\Omega_\ak^2)^2} \delta_{\ak \tilde \bk}\,,
\ee
one can express \eqref{eq:FS3_alldiff} in terms of the real mass parameters. The result is again \eqref{eq:FS3_stu}.

\section{Details about the universal multiplet} \label{app:universal_hyper}

We give here the details needed in order to describe the geometry of the $\frac{SU(2,1)}{SU(2)\times U(1)}$ manifold.
Its metric is given by
\be
ds^2 = %
2\left[  \frac{\abs{dz_1}^2 + \abs{dz_2}^2}{1 - \abs{z_1}^2 - \abs{z_2}^2}
+ \frac{\abs{ z_1^* dz_1 + z_2^* dz_2}^2}{\left( 1 - \abs{z_1}^2 - \abs{z_2}^2 \right)^2} \right] \,, \label{eq:SU21_metric}
\ee
where the overall normalization is chosen so that
\be
R=-12 %
\,.
\ee
This metric can be obtained by the following \Kahler potential
\be
\cK_H=%
-2\log \left(1-|z_1|^2-|z_2|^2\right) \label{eq:Kahler_hyp}\,.
\ee

The hypermultiplet scalars are the $z_i$ and $\bar z_i$ fields. To describe the manifold geometry, it's convenient to introduce a different set of coordinates $(\rho,\theta,\phi,\psi)$, related to the $z_i$'s as
\be
z_1 = \rho \cos \frac{\theta}{2} e^{i (\psi + \phi)/2} \,, \qquad
z_2 = \rho \sin \frac{\theta}{2} e^{i (\psi - \phi)/2} \,.
\ee
Introducing the frame
\ie
e^1 &= %
\frac{\rho}{\sqrt{2} \sqrt{1 - \rho^2} } \sigma_1 \,, \qquad \sigma_1 = \cos \psi d \theta + \sin \psi \sin \theta d\phi \,, \\
e^2 &= %
\frac{\rho}{\sqrt{2} \sqrt{1 - \rho^2} } \sigma_2 \,, \qquad \sigma_2 = -\sin \psi d \theta + \cos \psi \sin \theta d\phi \,, \\
e^3 &= %
\frac{\rho}{\sqrt{2} (1 - \rho^2)  } \sigma_3 \,, \qquad \sigma_3 =  d \psi +\cos \theta d\phi \,, \\ 
e^4 &= %
\sqrt{2} \frac{d\rho}{1 - \rho^2} \,,
\fe
then we can obtain the metric \eqref{eq:SU21_metric} in these new coordinates as $ds^2=e_a e^a=h_{uv}dq^u dq^v$.

We can also define some frame vectors ${f^{iA}}_u$. If we think of $f^{iA}$ as a one-form $f^{iA} = f^{iA}{}_u dq^u$, then
\be
f^{11} = \frac{i e^3 + e^4 }{\sqrt{2}}\,, \qquad f^{12} =\frac{ i e^1 - e^2 }{\sqrt{2}}\,, \qquad 
f^{21} = \frac{i e^1 + e^2}{\sqrt{2}} \,, \qquad f^{22} = \frac{- i e^3 + e^4}{\sqrt{2}} \,.
\ee
This choice frame vectors satisfies many properties, many of which we do not report here, but can be found for example in equation (4.142) of \cite{Lauria:2020rhc}; we mention only that
\ie
h_{uv}&={f^{iA}}_{u} \varepsilon_{ij} C_{AB} {f^{jB}}_v\\
\left( {f^{iA}}_{u}\right) ^*&= {f^{jB}}_{u}\varepsilon_{ji} \rho_{B \bar A}\,, \label{eq:f_properties}
\fe
with the choices
\be
C_{AB}= \varepsilon_{AB}, \qquad \rho_{A \bar B}=\varepsilon_{A \bar B}\,.
\ee
These two matrices are related by $C_{AB} = \rho_{A \bar C} {d^{\bar C}}_B$, so, in this case,
\be
{d^{\bar{A}}}_B = \delta^A_B \,,
\ee
such that $C_{AB} = \rho_{A \bar C} {d^{\bar C}}_B$. 
These frame fields define as well a hypercomplex structure
\be
\vec{J}\indices{_u^v}=-f\indices{^{iA}_u} f\indices{^v_{jA}} \vec{\tau}\indices{_i^j}\,,
\ee
where $ f\indices{^v_{jA}}$ is defined as the inverse of  $f\indices{^{jA}_v}$,
\be
f\indices{^{iA}_v} f\indices{^u_{iA}}=\delta^u_v\,, \qquad  f\indices{^{iA}_u} f\indices{^u_{jB}}=\delta_j^i \delta^A_B\,,
\ee
with $\vec{\tau}_i{}^j=i \vec{\sigma}_i{}^j$. This hypercomplex structures are covariantly constant up to a rotation:
\be
\nabla_w \vec{J}_u{}^v + 2 \vec{\omega}_w \times \vec{J}_u{}^v = 0 \,.
\ee	
In particular, an explicit calculation gives that the one-forms $\vec{\omega} = \vec{\omega}_w dq^w$ are
\be
\omega^1 = \frac{\sigma_1}{2 \sqrt{1 - \rho^2}} \,, \qquad
\omega^2 = \frac{\sigma_2}{2 \sqrt{1 - \rho^2}} \,, \qquad
\omega^3 = \frac{(2 - \rho^2) \sigma_3 }{4 (1 -\rho^2)} \,.
\ee
The manifold $\frac{SU(2,1)}{SU(2)\times U(1)}$ has eight Killing vectors (see equation (12) of\cite{Behrndt:2000ph}). We are interested in two of them,
\be
\zeta = \partial_\phi = \frac{i( z_1 \partial_{z_1} - z_2 \partial_{z_2} -\bar z_1 \partial_{\bar z_1} + \bar z_2 \partial_{\bar z_2} ) }{2}    \,, \qquad \xi = \partial_\psi = \frac{i( z_1 \partial_{z_1} + z_2 \partial_{z_2} -\bar z_1 \partial_{\bar z_1} - \bar z_2 \partial_{\bar z_2} ) }{2}  \,. \label{eq:Kill_vec_basis}
\ee 

It can be checked that they obey $\nabla_{(u} \zeta_{v)} = \nabla_{(u} \xi_{v)} =0$.  From the Killing vectors, we can obtain the moment maps
\ie
P_{(\zeta)} &= %
- \frac{1}{2} \vec{J}_u{}^v \nabla_v \zeta^u
= -%
\begin{pmatrix}
	\frac{\sin \theta \sin \psi}{\sqrt{1 - \rho^2}}, & \frac{\sin \theta \cos \psi}{\sqrt{1 - \rho^2}}, & \frac{(2 - \rho^2) \cos \theta}{2(1 - \rho^2)} 
\end{pmatrix} \,, \\
P_{(\xi)} &=   %
-\frac{1}{2}\vec{J}_u{}^v \nabla_v \xi^u
= -%
\begin{pmatrix} 
	0, & 0, & \frac{\rho^2}{2(1 - \rho^2)} 
\end{pmatrix} \,, \label{eq:moment_maps}
\fe
which are related to the Killing vectors by
\be
\p_u \vec{P}_I+2\vec{\omega}_u \times \vec{P}_I=\vec{J}_{uv} k_I^v\,,
\ee
and
\be
-2%
n_H \vec{P}_I=\vec{J}_u{}^v \nabla_v k_I^u\,.
\ee

\bibliographystyle{ssg}
\bibliography{S3}

\end{document}